\documentclass[conference]{IEEEtran}
\usepackage{hyperref}
\usepackage{setspace}

\usepackage{cite}
\usepackage[caption=false,font=footnotesize]{subfig}
\usepackage{url}

\newcommand{\dIg}{\hbox{${S}$}}
\newcommand{\dSp}{\hbox{$I$}}
\newcommand{\dSt}{\hbox{$R$}}

\newcommand{\Ig}{\hbox{$\mathcal{S}$}}
\newcommand{\Sp}{\hbox{$\mathcal{I}$}} 
\newcommand{\St}{\hbox{$\mathcal{R}$}} 
\newcommand{\SIR}{\dIg\dSp\dSt}

\newcommand{\igual}{\hbox{ $\leftarrow$ }}

\usepackage{colortbl}
\usepackage{color}
\usepackage{graphicx}
\usepackage{epstopdf}
\usepackage{amsthm}
\usepackage{algorithm,algorithmic}
\usepackage{amsmath}
\usepackage{balance}

\definecolor{Gray}{gray}{0.87}

\theoremstyle{definition}
\newtheorem*{definition}{Definition}

\begin{document}


%
\title{The Impact of Social Curiosity on Information Spreading on Networks}



%
\author{\IEEEauthorblockN{Didier A. Vega-Oliveros\IEEEauthorrefmark{1}\IEEEauthorrefmark{3},
Lilian Berton\IEEEauthorrefmark{2},
Federico Vazquez\IEEEauthorrefmark{3}, and
Francisco A. Rodrigues\IEEEauthorrefmark{1}}
\IEEEauthorblockA{\IEEEauthorrefmark{1}Instituto de Ci\^{e}ncias Matem\'{a}ticas e de Computa\c{c}\~{a}o,\\
Universidade de S\~{a}o Paulo - Campus de S\~{a}o Carlos,\\
PB 668, Zip code 13560-970 S\~{a}o Carlos, SP, Brazil. \\
Email: {davo, francisco}@icmc.usp.br}
\IEEEauthorblockA{\IEEEauthorrefmark{2}
Universidade Federal de S\~{a}o Paulo, Campus de S\~{a}o Jos\'{e} dos Campos\\
Zip code 12247-014 S\~{a}o Jos\'{e} dos Campos, SP, Brazil.}
\IEEEauthorblockA{\IEEEauthorrefmark{3}Instituto de F\'{i}sica de L\'{i}quidos y Sistemas Biol\'{o}gicos, IFLYSIB, \\
Universidad Nacional de la Plata - 1900 La Plata, Argentina.}}


\maketitle

\begin{abstract}
Most information spreading models consider that all individuals are identical psychologically. They ignore, for instance, the curiosity level of people, which may indicate that they can be influenced to seek for information given their interest. For example, the game Pok{\'{e}}mon GO spread rapidly because of the aroused curiosity among users. This paper proposes an information propagation model considering the curiosity level of each individual, which is a dynamical parameter that evolves over time. We evaluate the efficiency of our model in contrast to traditional information propagation models, like $SIR$ or $IC$, and perform analysis on different types of artificial and real-world networks, like Google+, Facebook, and the United States roads map. We present a mean-field approach that reproduces with a good accuracy the evolution of macroscopic quantities, such as the density of stiflers, for the system's behavior with the curiosity. We also obtain an analytical solution of the mean-field equations that allows to predicts a transition from a phase where the information remains confined to a small number of users to a phase where it spreads over a large fraction of the population.  The results indicate that the curiosity increases the information spreading in all networks as compared with the spreading without curiosity, and that this increase is larger in spatial networks than in social networks. When the curiosity is taken into account, the maximum number of informed individuals is reached close to the transition point.  Since curious people are more open to a new product, concepts, and ideas, this is an important factor to be considered in propagation modeling. Our results contribute to the understanding of the interplay between diffusion process and dynamical heterogeneous transmission in social networks.
\end{abstract}

%
\IEEEpeerreviewmaketitle

\section{Introduction}

Information spreading research relates to understanding how rumors, news and behaviors spread on a large scale in a short time. The diffusion of information is a ubiquitous process in the society, where people are interacting with each other all the time, making contact and exchanging information in the public transport, at work or school, the Internet and at home~\cite{Noe2016,Tabacchi2017}. In this way, the maximization of the transmission of some information on social networks and the society is important for scenarios like viral marketing, political and ideological campaigns, among others.

After the establishment of Network Sciences~\cite{barabasiEalbert1999,newman2010networks}, more accurate models have been introduced for analyzing the spread of information as social contagion processes~\cite{Guille2013,Pastor-Satorras2015,Kempe15,porter2016,Vega-OliverosH17}. Within these models, the $\SIR$ epidemic model has been applied extensively in research areas like opinion formation, decision making, and information or rumor propagation~\cite{Guille2013,Pastor-Satorras2015}. The spreading of information can be approached as a psychological contagion where an idea passes from person to person and  ``contaminates'' the mind of many individuals~\cite{Vega-OliverosH17}. In the classical approach people are divided into three groups or states: ignorant (those not aware of the information), spreaders (those who spread the information), and stiflers (those who know the information but stop trying to convince others).  



However, most models on spreading processes consider homogeneous or static transmission probabilities~\cite{newman2010networks, Guille2013,Pastor-Satorras2015}, i.e., each individual has a fixed likelihood of transmission over time. Moreover, the curiosity has been disregarded in these models until now, although it is expected to have an impact on diffusion processes: the more surprising a rumor, the more rapidly and extensively it is likely to spread.  Also, the more people know the information and the more sources that spread it, the higher the likelihood of accepting that information. The curiosity is psychologically defined as the desire of knowledge and experiences that lead to exploratory behavior and the acquisition of new information~\cite{ZUCKERMAN:1986,Litman:2005,Wu:2013}. It is also associated with the reward of two kind of triggers, the cognitive curiosity and sensory curiosity~\cite{ZUCKERMAN:1986}. A person often gets curious about some behaviors of his/her friends; a phenomenon known as the social curiosity in psychology~\cite{Wu:2013}.

\begin{figure}[!bt]
\centering	
  \subfloat[]{\label{fig:1a}\includegraphics[width= 0.17\textwidth]{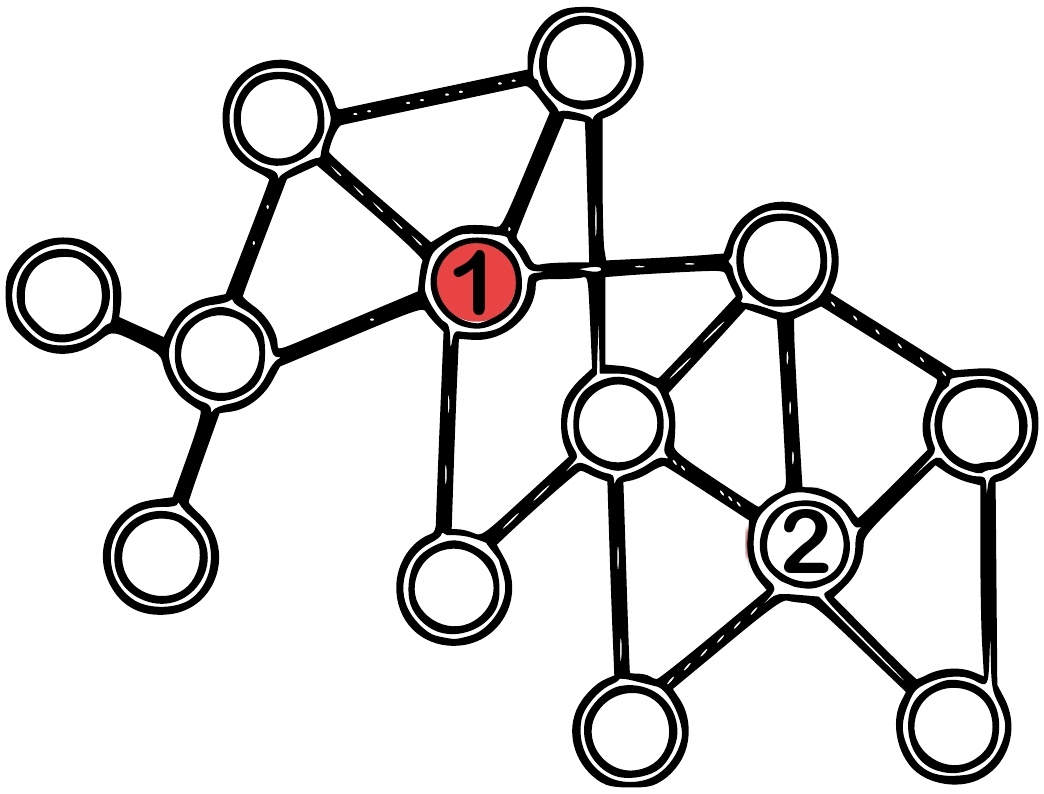}} \hspace{0.3cm}  
  \subfloat[]{\label{fig:1b}\includegraphics[width= 0.17\textwidth]{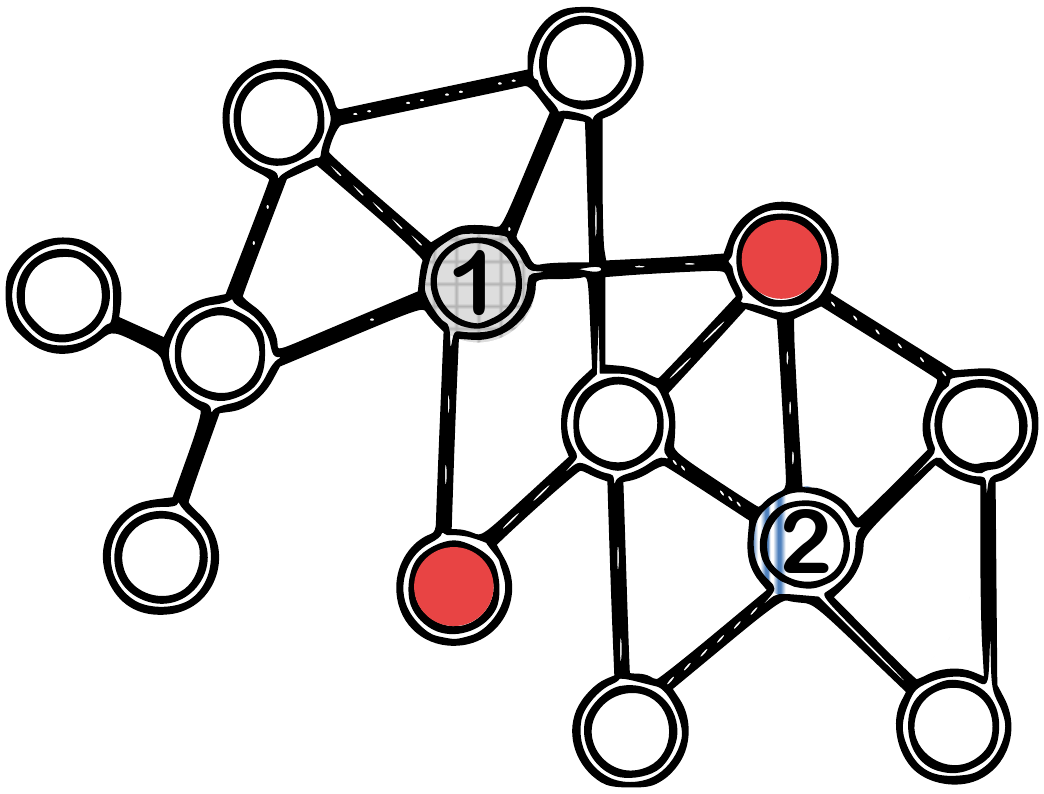}}\vspace{-0.3cm}   
  
  \subfloat[]{\label{fig:1c}\includegraphics[width= 0.17\textwidth]{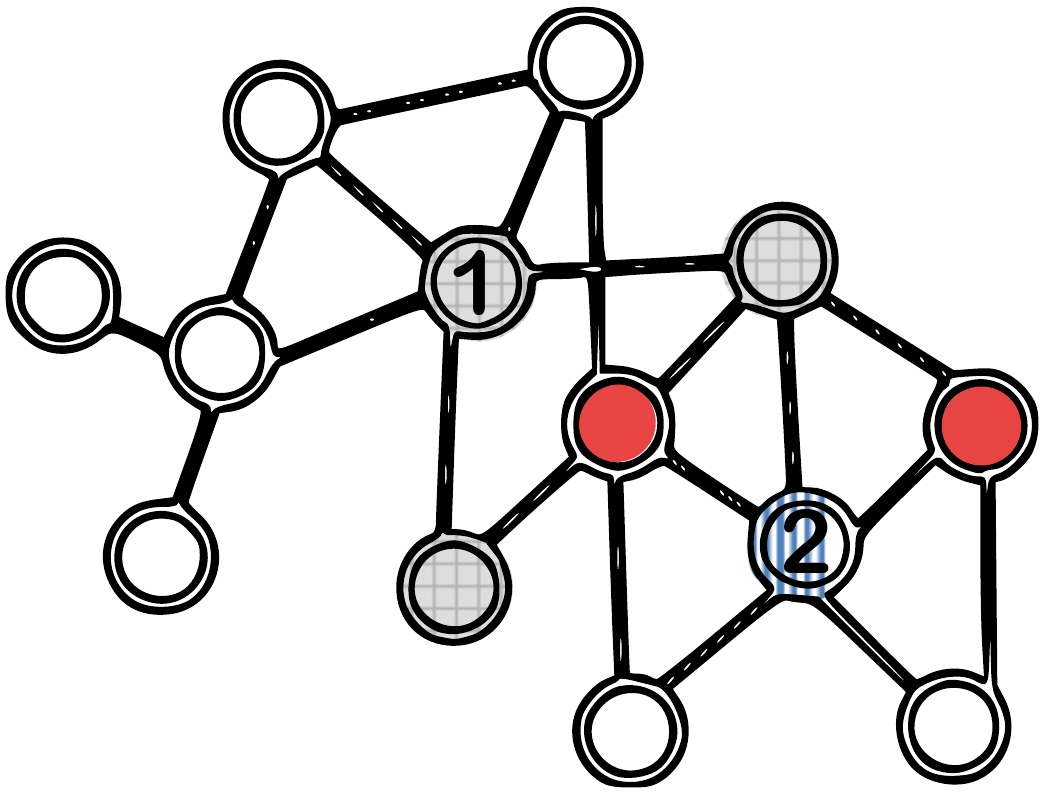}} \hspace{0.3cm}
  \subfloat[]{\label{fig:1d}\includegraphics[width= 0.17\textwidth]{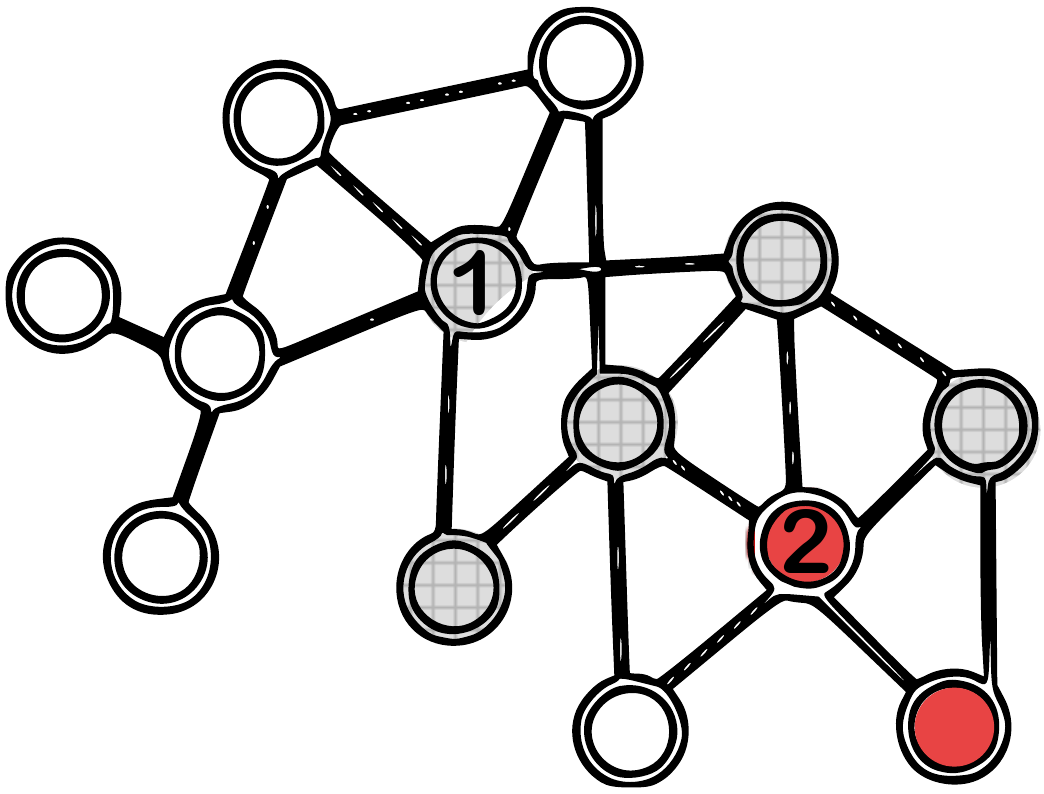}}   

\caption{\label{fig:exemplo} 
 (Color online) Illustration of four consecutive time steps in the Social Curiosity model for information spreading on networks. (a) Vertex \emph{1} is the only initial spreader (red circle).  Vertex \emph{2} is ignorant (white circle) and has no curiosity because none of its neighbors knows about the information. (b) Vertex \emph{1} randomly convinces two neighbors and it turns into stifler (gray circle). Vertex \emph{2} starts to become curious about the information (blue bars inside the circle) since one of its five neighbors is informed. (c) Given that already three neighbors of vertex \emph{2} know the information, its curiosity level and thus the chances to get the information are increased. (d) Eventually, vertex \emph{2} becomes informed due to its curiosity and the contact with spreaders.}\vspace{-0.3cm}
\end{figure}


In this paper, we aim to model the information spreading on networks taking into account the human curiosity. We evaluated how information (for instance a rumor) propagates considering the social curiosity, which is high if more friends are talking about the spreading rumor (Figure \ref{fig:exemplo}). 
One motivational example is the rapid outbreak of the game Pok{\'{e}}mon GO \cite{NYT:2016,Tabacchi2017}  launched by Niantic in July 2016 and that is based on the 1990's Pok{\'{e}}mon video games. It became a social phenomenon that achieved a high popularity in a short time, being the app in the iOS App Store with more downloads in one week than any other in the same period, and one of the most commented topics in on-line and traditional magazines, Social Networks, among others~\cite{Tabacchi2017}. This success is probably due to the fact that the game promotes an interesting mix of virtual reality and socialization activities in public spaces where players can interact, which arouses the curiosity of people that are not part of the game~\cite{Wu:2013}.  The probability that a person becomes player increases if more of its neighbors or friends are playing it.  In this article, we show by analytical and numerical simulations that including the psychological characteristics of individuals, such as the curiosity, improves the diffusion of information in traditional spreading models.

The major contributions of this work are: \textit{i}) we provide a more detailed and realistic description of information spreading processes with the combination of the curiosity mechanism and the {\SIR} model of epidemics. \textit{ii}) We define a dynamical measure of the social curiosity for each subject in its neighborhood. \textit{iii}) A mathematical model for information spreading is presented, including different probabilities of contagion and the dynamical curiosity that evolves over time. \textit{iv}) We provide an analytical solution for the final density of informed individuals, which is $\mathcal{O}(1)$ and agrees quite well with simulations results.  \textit{v}) We analyze the impact of curiosity on information spreading, such as the critical threshold, the peak of spreaders and the final density of informed individuals, in artificial and real-world networks. The results indicate that depending on the transmission parameters, the social curiosity improves the propagation process in nearly twice the expected value of the non-curiosity case.  Besides, the curiosity increases the peak in the number of spreaders reached during the propagation process and speeds up the propagation.  These results suggest that the social curiosity is an important feature to be considered in spreading processes, by marketing campaigns and in the analysis of information spreading models.

The remaining of this paper is organized as follows. Section \ref{related-work} presents the literature on related models for information diffusion on networks. Section \ref{concepts} brings some concepts and definitions used in the paper.  Section \ref{proposal} introduces the proposed model for information spreading that includes the social curiosity mechanism. Section \ref{experiments} shows some experimental results in artificial and real-world networks.  In Section~\ref{mean-field} we develop a mean-field approach for the model's dynamics and obtain analytic results. Finally, Section \ref{conclusions} provides the final remarks and future work.

\section{Related works} \label{related-work}

Several models have been proposed for modeling propagation dynamics on networks~\cite{Guille2013, Pastor-Satorras2015,Kempe15,porter2016,Vega-OliverosH17,Zhang2016,Zhao2011,Nekovee2007,Vega-Oliveros-socinf15}. These models consider some assumptions about the propagation process and network structure, like the degree correlation or distribution, classes of vertices, among others~\cite{Guille2013,Pastor-Satorras2015,Kempe15,porter2016}. The majority of the spreading models consider only homogeneous or fixed transmission probabilities, i.e., each vertex has the same likelihood of transmitting the information~\cite{Kempe15,Pastor-Satorras2015,porter2016}. For instance, in \cite{Vega-OliverosH17} the authors propose a discrete-time model for rumor propagation with heterogeneous transmission probabilities, but still, each vertex has a constant probability over time. Other transmission models consider characteristics like short-term of immunity or steady active state~\cite{newman2010networks}, and spreaders' procrastination in the transmission of information by heterogeneous delays~\cite{Zhang2016}. Also, there are some proposals that introduced disbelieving or forgetting mechanism~\cite{Zhao2011}, lost of interest~\cite{Nekovee2007}, apathy and many others~\cite{Pastor-Satorras2015}. Regarding the contagion dynamic, the diffusion process can be spontaneous or by different kind of contact interaction or adaptation among the subjects~\cite{Vega-Oliveros-socinf15}. 

Previous studies analyzed the effect of the new characteristic in the propagation process and examined numerical and dynamical properties on top of the different type of networks. They looked for critical thresholds, computational cost and accuracy in the results~\cite{Pastor-Satorras2015}. However, these studies of information spreading assumed fixed parameters and the curiosity mechanism was also disregarded~\cite{Guille2013,Kempe15,Pastor-Satorras2015,porter2016}. Here, we consider the curiosity as an important factor to determinate the speed and coverage of information spreading on networks.

\section{Concepts and methods} \label{concepts}

An \emph{information system} can be represented as a network (or a graph) consisting of two elements: actors (individuals or objects that are the elements of the system) and the connections (interactions, relationships, or social ties)~\cite{newman2010networks}. Formally, let's consider a network $G = (V, E)$, where $V$ is the set of $|V|= N$ vertices, $E$ is the set of $|E| = M$ edges or connections. The edges can be directed, indicating the flow of the relation, or undirected. 
Two vertices are called \emph{neighbors} if they are connected by an edge.

Classical propagation models consider that nodes interact in the same way during the dynamic. The model's parameter $\beta$ (for propagating or transmitting the signal) and $\mu$ (for recovering or stopping the propagation) are constants and invariant during the process. In the $\SIR$ model~\cite{newman2010networks,Pastor-Satorras2015}, the susceptibles (\dIg) are those who remain unaware of the information, the spreaders (\dSp) are the influencers who disseminate the information, and the stiflers (\dSt) are those who know the information but lose the interest in the spreading process. The influencers and the stiflers are part of the group of informed individuals. Also, the transition to the stifler state happens spontaneously~\cite{Pastor-Satorras2015}. Whenever a spreader $i$ meets a susceptible neighbor $j$, the latter become spreader with probability $\beta$. Spreaders turn into stifler with probability $\mu$.

The effective spreading rate~\cite{Hethcote2006,Buono2013,Pastor-Satorras2015} $\lambda$ = $\beta / \mu$ is the contagion strength or infectivity level of the propagation. The final density of informed individuals is equivalent to the combination of $\beta$ and $\mu$ that hold the same $\lambda$ value~\cite{newman2010networks,Hethcote2006,Vega-OliverosH17}. In epidemics, this is known as the basic reproductive ratio\cite{newman2010networks} and a global endemic phase is reached by infectivity levels above a critical threshold ($\lambda_c$), while no endemic phase appear whenever the contagion strength is below $\lambda_c$. For random and uncorrelated networks, $\lambda_c=\langle k \rangle / \langle k^2 \rangle$~\cite{Pastor-Satorras2015}, where $\langle k \rangle$ and $\langle k^2 \rangle$ are the mean and the second moment of the degree distribution, respectively.

Another propagation model is the Independent Cascade ($IC$) model~\cite{Kempe15,porter2016}. The difference with the $\SIR$ model yields in the transition to the stifler (inactive spreaders) state~\cite{Guille2013}, which happens instantly in the next iteration step. In fact, the $IC$ can be approached as a variation of the {\SIR} model~\cite{Vega-OliverosPRE2017}. Both models are spreader-centric, i.e., the contagion occurs given the contact interaction of the spreaders. In the $IC$ model, the spreaders try to pass the information only once, which is equivalent to the {\SIR} with parameter $\mu = 1$. The models are also incremental monotonic functions, and they can consider heterogeneous probabilities in the diffusion~\cite{Buono2013,porter2016,Vega-OliverosH17}.

\section{Curiosity in spreading process} \label{proposal}

We present a discrete-time approach including the social curiosity property, centered in the propagation dynamic of each vertex on arbitrary network structure. The approach follows the $\SIR$ states and the social curiosity is as a parameter that evolves over time. This parameter introduces a susceptible-centric approach, affecting the behavior of each susceptible vertex. We consider vertices with different curiosity levels and this approach is rooted on the observation that activity patterns are essentially heterogeneous~\cite{Vega-OliverosH17} and people do not behave in the same way when asking for information.


 \begin{figure}
     \centering
     \includegraphics[width= 0.19\textwidth]{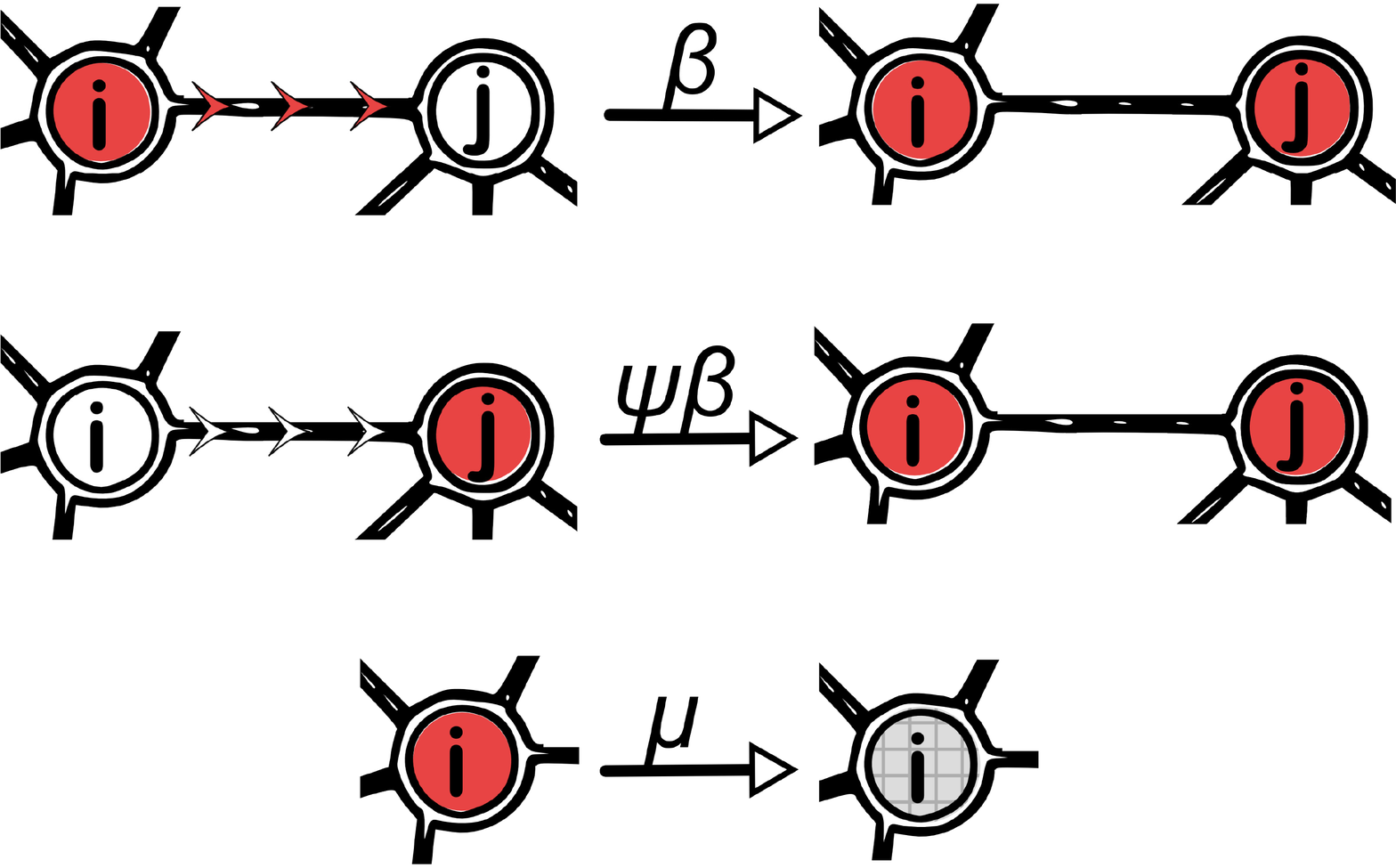}\vspace{-0.3cm}
     \caption{(Color online) Dynamical rules of the proposed model. Vertex $i$ is making contact with the neighbor $j$, in which: the first line, $i$ is a spreader (red circles) that convinces its ignorant neighbor (white circles) with probability $\beta$; the second line, $i$ is an ignorant that asks for information to its spreader neighbor with probability $\Psi$ and get convinced with probability $\beta$. The third line is the spontaneous transition of $i$ to stop the propagation of the information (stifler state in gray circles). }\vspace{-0.3cm}
     \label{eq:curiosityRules}
 \end{figure}
 

\subsection{Curiosity measurement}

The personality, as an inherent characteristic of an individual, has a role in the organization of people on the network, and in turn, they position reinforce specific personality traits~\cite{Noe2016}. A range of network-based features is correlated with specific characteristics of individuals~\cite{viswanath09,Noe2016,Tabacchi2017}. For instances, the Extroversion, one of the Five-Factor Model of personality~\cite{Noe2016,Tabacchi2017}, is positively correlated with the number of friends or connections of a person~\cite{Noe2016}. In particular, the Openness to experience refers to an individual's curiosity and willingness to engage in new experiences. The Extroversion and Openness are positively related with the capacity of higher social engagement~\cite{Noe2016}. The before is in accordance with a social study about the users of Pok{\'{e}}mon GO~\cite{Tabacchi2017}, where was reported that individuals with high level of Openness also were the early adopters.

In the case of epidemic or information spreading process, the ``early adopters'' are correlated to the network centrality, where the most central nodes are reached at an early stage~\cite{Vega-OliverosPRE2017}. This is explained by the higher interaction of individuals with their pairs and the update of their states as a consequence of these interactions. This property is known as the homophily phenomena or assortative mixing~\cite{Vega-OliverosPRE2017}, the tendency of similar nodes to be connected, which is the main characteristic of social relationship. 

\begin{definition}{(\textit{Curiosity Strength})}
The curiosity strength $\Psi$ is the probability that a susceptible vertex asks a spreader neighbor about some information.
\end{definition}

Thus, given that the Openness to experience individuals are part of the group of early adopters and they are somehow positively correlated with the connection structure of the network, we can consider that higher the scores in Openness to experience, the more strongly connected the individual.
For a more realistic scenario, we define the social curiosity for a given vertex ($\Psi_i(t)$) as the fraction of informed neighbors over time (Figure~\ref{fig:exemplo}). Thus, the higher the level of social curiosity, the higher the probability of a susceptible vertex makes contact with a spreader. Therefore, the chances of the susceptible vertices take an action for searching for information depends on the propagation process and the size of their networks.

When $\Psi = 0$, we recover the {\SIR} and the $IC$ models. The Utopian case happens when all the individuals are always $100\%$ curious ($\Psi = 1$). The critical threshold for the maximum curiosity case, $\lambda_c^{1}$, and the critical threshold of non-curiosity case, $\lambda_c^{0}$, are the lower and upper bound of the critical threshold for the proposed social curiosity measure ($\Psi_i(t)$), i.e, $\lambda_c^{1} < \  \lambda_c^{\Psi_i(t)} \ \leq \lambda_c^{0}$.


\subsection{Monte Carlo simulation}


We consider a constant population of $N$ vertices in all time steps. The spreading dynamic is a stochastic process where each vertex has a probability of being in a possible state over time. The density of susceptible $\dIg(t)$, spreader $\dSp(t)$ or stifler $\dSt(t)$ vertices satisfies  $\dIg(t) \: + \: \dSp(t) \: + \: \dSt(t) \: = \: 1$ for all time steps. At each time step, all spreaders uniformly try to infect their neighbors with probability $\beta$, or stop the diffusion with probability $\mu$. 
The dynamical rules that define the general diffusion process are depicted in Figure~\ref{eq:curiosityRules}.

\begin{center}
\begin{algorithm}[H]
\scalebox{0.8}{
\begin{minipage}{\linewidth}
\begin{algorithmic}[1]
\caption{MC for information spreading with curiosity.}
\label{alg:MC}
\REQUIRE $G$,  $\Ig(0)$,  $\Sp(0)$,  $\St(0)$, $\beta$, $\mu$, $\Psi$
\\ \textit{Initialization} : $t \igual 0$
\REPEAT
\STATE $t \igual t + 1$
\STATE $\Ig(t) \igual \Ig(t-1) $, $\Sp(t) \igual \Sp(t-1)$, $\St(t) \igual \St(t-1)$
\FORALL{vertex $i$ such that $i \in \Sp(t-1)$ }
    \FORALL{neighbor $j$ of $i$ such that $j \in \Ig(t-1)$ }
        \IF {$i$ infected $j$ with probability $\beta$} 
            \STATE $\Ig(t) \igual \Ig(t) - \{j\}$ 
            \STATE $\Sp(t) \igual \Sp(t) \bigcup \{j\}$ \COMMENT {update the states}            
        \ENDIF    
    \ENDFOR
    \IF {$i$ gets recovered with probability $\mu$} 
        \STATE $\Sp(t) \igual \Sp(t) - \{i\}$
        \STATE $\St(t) \igual \St(t) \bigcup \{i\}$ 
    \ENDIF
\ENDFOR
\FORALL{vertex $i$ such that $i \in \Ig(t-1)$ }    
    \FORALL{neighbor $j$ of $i$ such that $j \in \Sp(t-1)$ }
        \IF{$i$ gets curious with probability $\Psi(i)$ \AND \\
            $i$ gets infected by $j$ with probability $\beta$}            
                \STATE $\Ig(t) \igual \Ig(t) - \{i\}$ 
                \STATE $\Sp(t) \igual \Sp(t) \bigcup \{i\}$                
            \ENDIF                    
    \ENDFOR    
\ENDFOR
\UNTIL{$\dSp(t) = 0$}
\RETURN $\dSt(t)$
\end{algorithmic}
\end{minipage}
}
\end{algorithm}
\end{center}

The Monte Carlo (MC) algorithm for the simulations consider the following variables at time $t = 0$:  (i) $\Ig(0)$ that represents the set of ignorants, (ii) $\Sp(0)$ the set of spreaders or seeds and (iii) $\St(0)$ the set of stiflers. Vertices belong to only one of the states over time, i.e., if $i \in \Sp(t) \: \hbox{, then }  \: i \notin [\Ig(t) \: \bigcup \: \St(t)]$, and always  $i \in [\Ig(t) \: \bigcup \: \Sp(t)\: \bigcup \: \St(t)]$. In terms of MC implementation, we assume that infection and recovering do not occur during the same time step. The contact interaction happens with all the neighbors of an informed node in each time step, known as reactive process (RP)~\cite{Vega-OliverosB15}. The macroscopic densities of susceptible ($\dIg(t)$) over time is calculated as $\dIg(t) = {|\Ig(t)|}/N$, where $|\Ig(t)|$ is the size of the set. For the other states, we adopt a similar approach.

The end of the simulation is reached when $\Sp(\infty) = \{ \: \emptyset \: \}$. Thus, the final fraction of informed individuals is $\dSt(\infty) = {|\St(\infty)|}/{N}$ or $\dSt(\infty) =  1 - {|\Ig(\infty)|}/{N}$. 
The algorithm for the information spreading with curiosity (Algorithm~\ref{alg:MC}) needs as input the network $G$, the set of initial states of the vertices $\Ig(0)$,  $\Sp(0)$,  $\St(0)$, the transmission probabilities $\beta$, $\mu$ and the vector $\Psi$ of node curiosity. The sets of vertices states can be approached as a structured list of nodes in a specific time step. For illustrative purpose, the steps (7 - 8), (12 - 13) and (19 - 20) are the same procedure for updating the state of the vertex, but it can be generalized in a separated function. Also, the part of propagation by curiosity, steps (16 - 23), can be refactored in a main loop (step 4) that considers all the vertices.

The complexity of the MC algorithm is determined by: (i) the maximum number of steps $T$ in which $\dSp(T) = 0$; (ii) the number of iterations in the loops of the classical (step 4) and curiosity (step 16) propagation, which are  $|\Ig(t) \bigcup \Sp(t)|$ and can be approximated to $N/2$; (iii) the computational cost of copying and updating the sets, which can be considered as a constant value $C_1$; and (iv) the cost of contacting the neighbors of the vertices, which in average for the whole networks is equivalent to $\langle k \rangle$. Thus, the complexity of the simulation algorithm is $\mathcal{O}(T * N/2 * C_1*(2\langle k \rangle + 1 ) ) \equiv \mathcal{O}(T * N * \langle k \rangle) $.

\section{Experiments} \label{experiments}

\begin{table}[!bt]
	\centering
	\caption{Topological properties of the networks considered here.}
	\resizebox{0.50\textwidth}{!}{
    \begin{tabular}{ c c c c c c c }
	\hline\
 	& Network & $N$ & $\left\langle k \right\rangle$ & $\left\langle k^2 \right\rangle$ & $\left\langle \ell \right\rangle$ & $\rho $ \\ \hline
 & ER & $1000$ & $11.9$ & $155$ &  $3.03$ & $0.001$   \\	
Artificial	& BA & $1000$ & $11.9$ &  $284$ &  $2.84$ & $-0.032$   \\		
	& SSF& $1000$ & $11.9$ &  $243$ & $4.075$  & $0.19$  \\	
	\hline		
	&\emph{USAroad} &  $6443$  &  $3.09$ & $10.41$ & $50.84$ & $0.19$ \\
Real &\emph{Google+} & $23613$ & $3.32$ &  $1251.67$ & $4.03$ & $-0.389$  \\
	&\emph{Facebook} & $63392$ & $25.77$ & $2256.80$ &  $4.32$ & $0.177$ \\	
	\hline
	\end{tabular}
     }
	\label{tab:DescricaoDasBasesUtilizadas}
\end{table}

\begin{figure*}[!tb]
\centering	
  \subfloat[BA]{\label{fig:modelBA}\includegraphics[width= 0.3\textwidth]{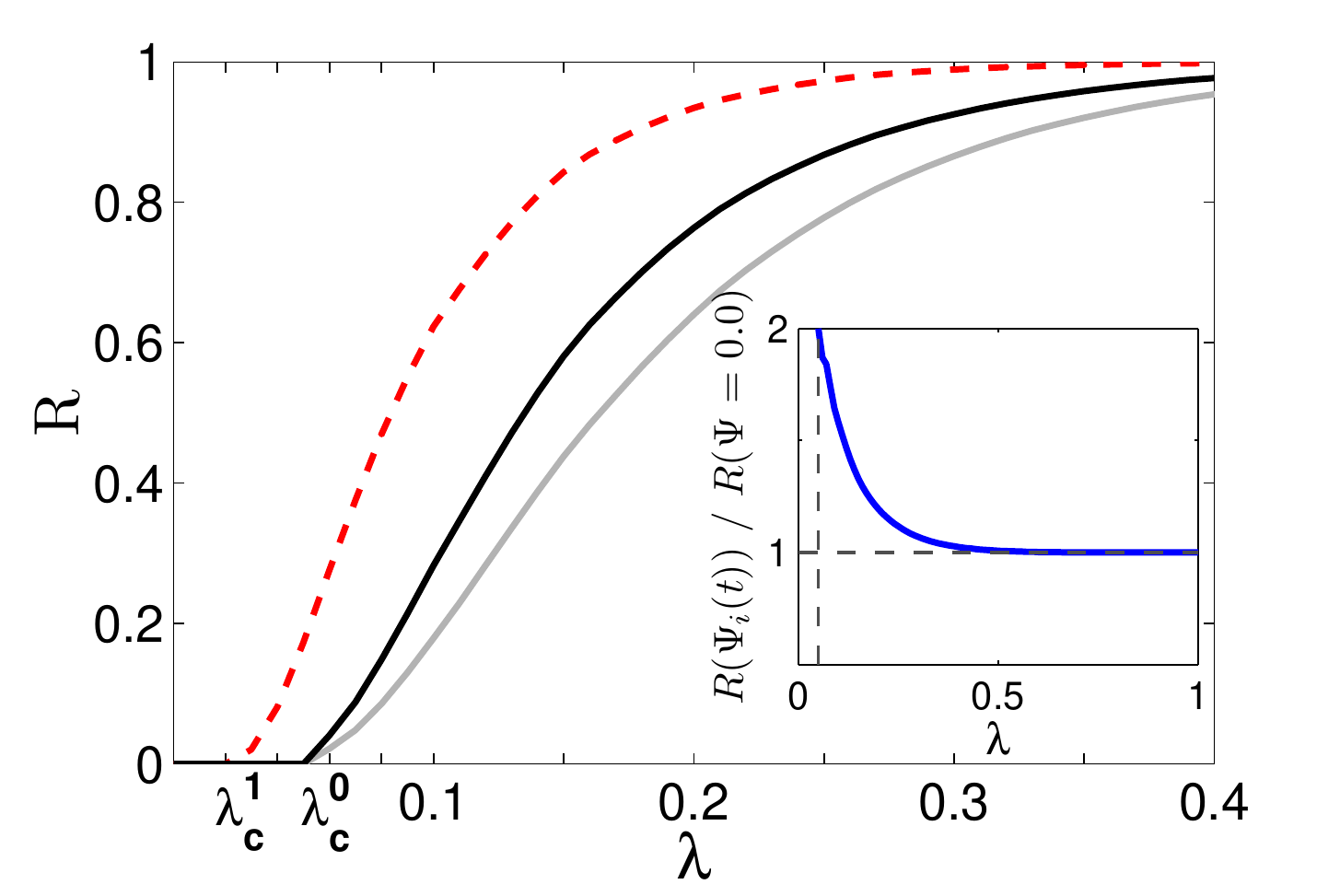}}  
\subfloat[ER]{\label{fig:modelER}\includegraphics[width= 0.3\textwidth]{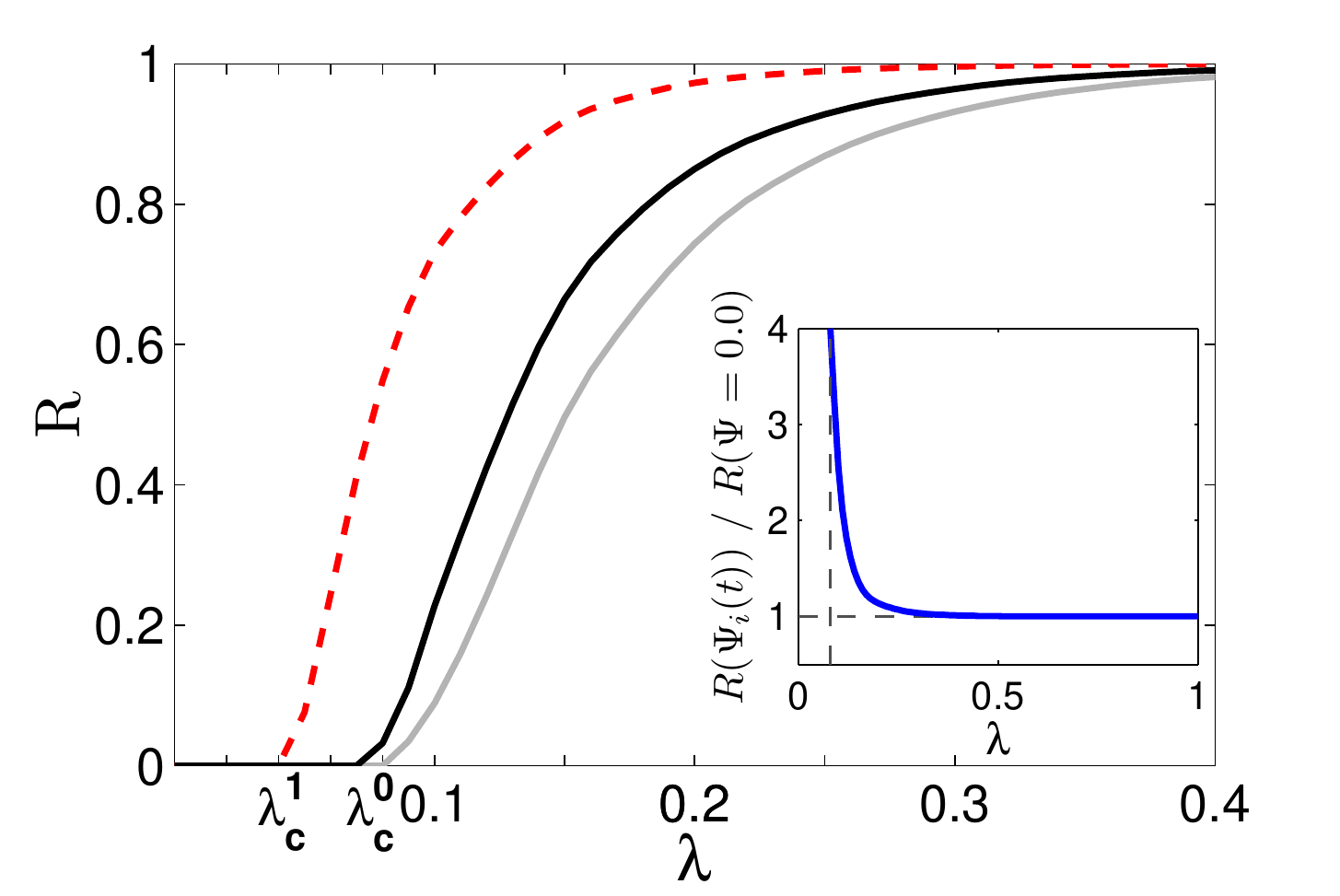}}   
  \subfloat[SSF]{\label{fig:modelSSF}\includegraphics[width= 0.3\textwidth]{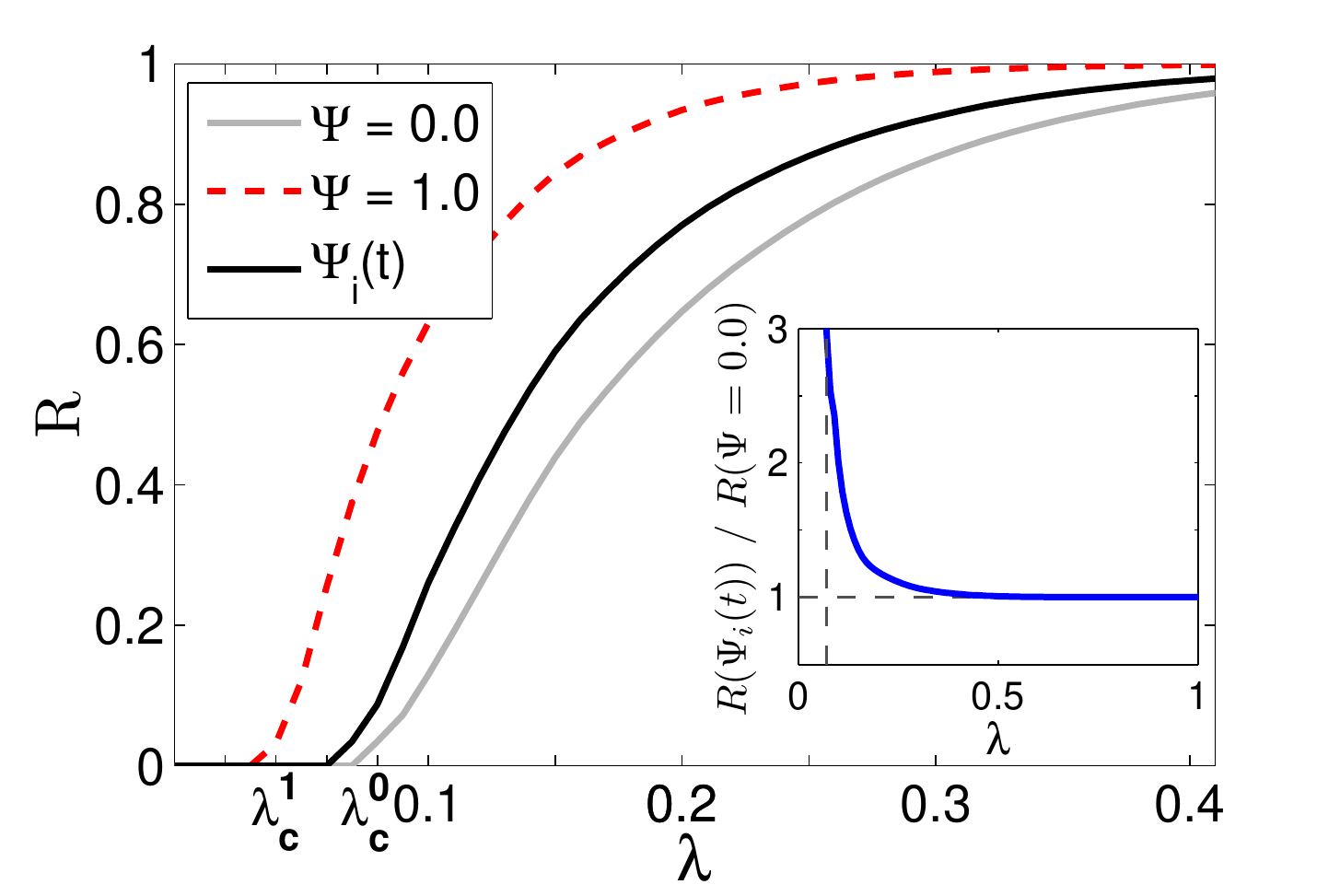}}\vspace{-0.3cm}   
  
  \subfloat[\emph{Google+}]{\label{fig:GPLUS}\includegraphics[width= 0.3\textwidth]{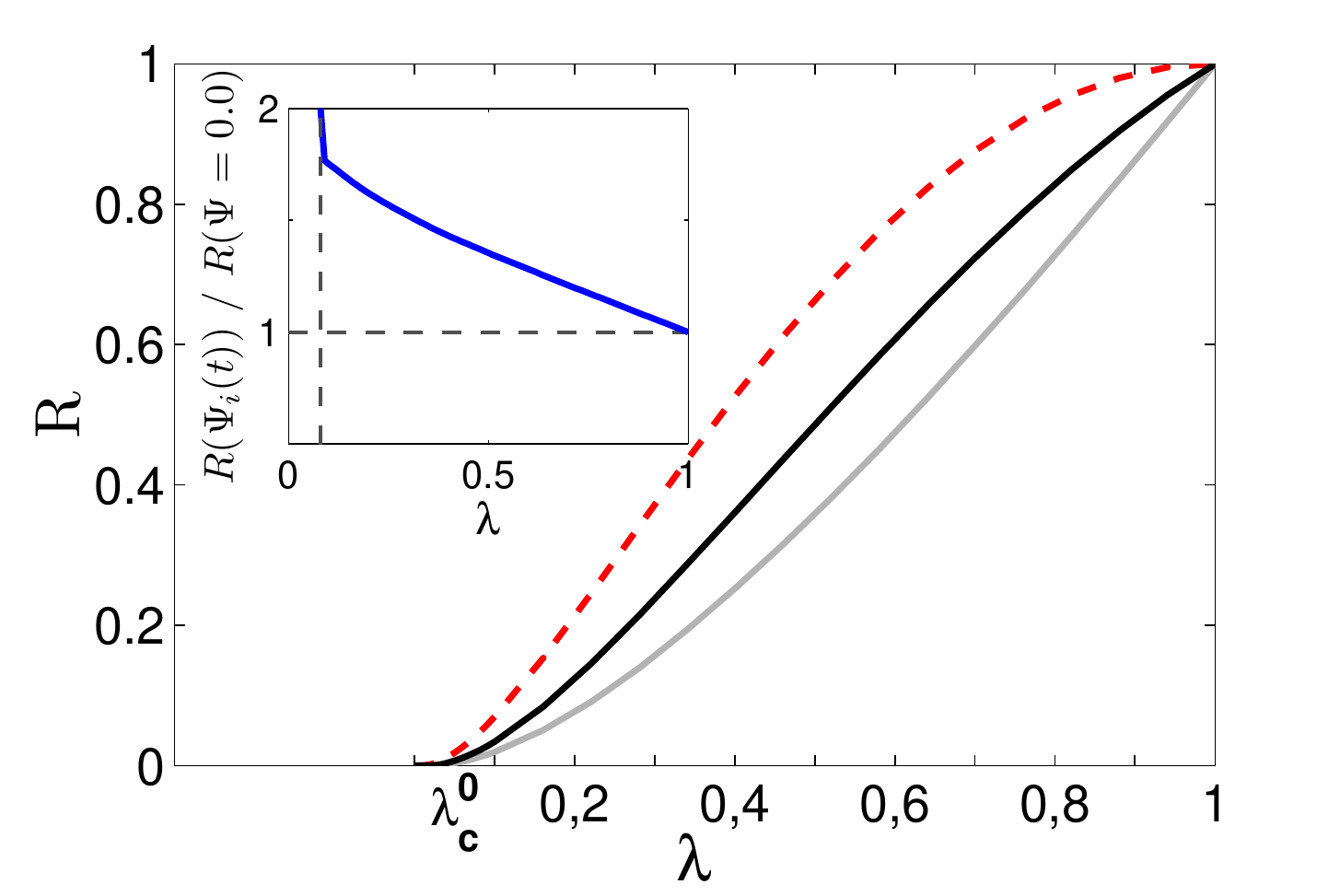}}    
\subfloat[\emph{USAroad}]{\label{fig:USA}\includegraphics[width= 0.3\textwidth]{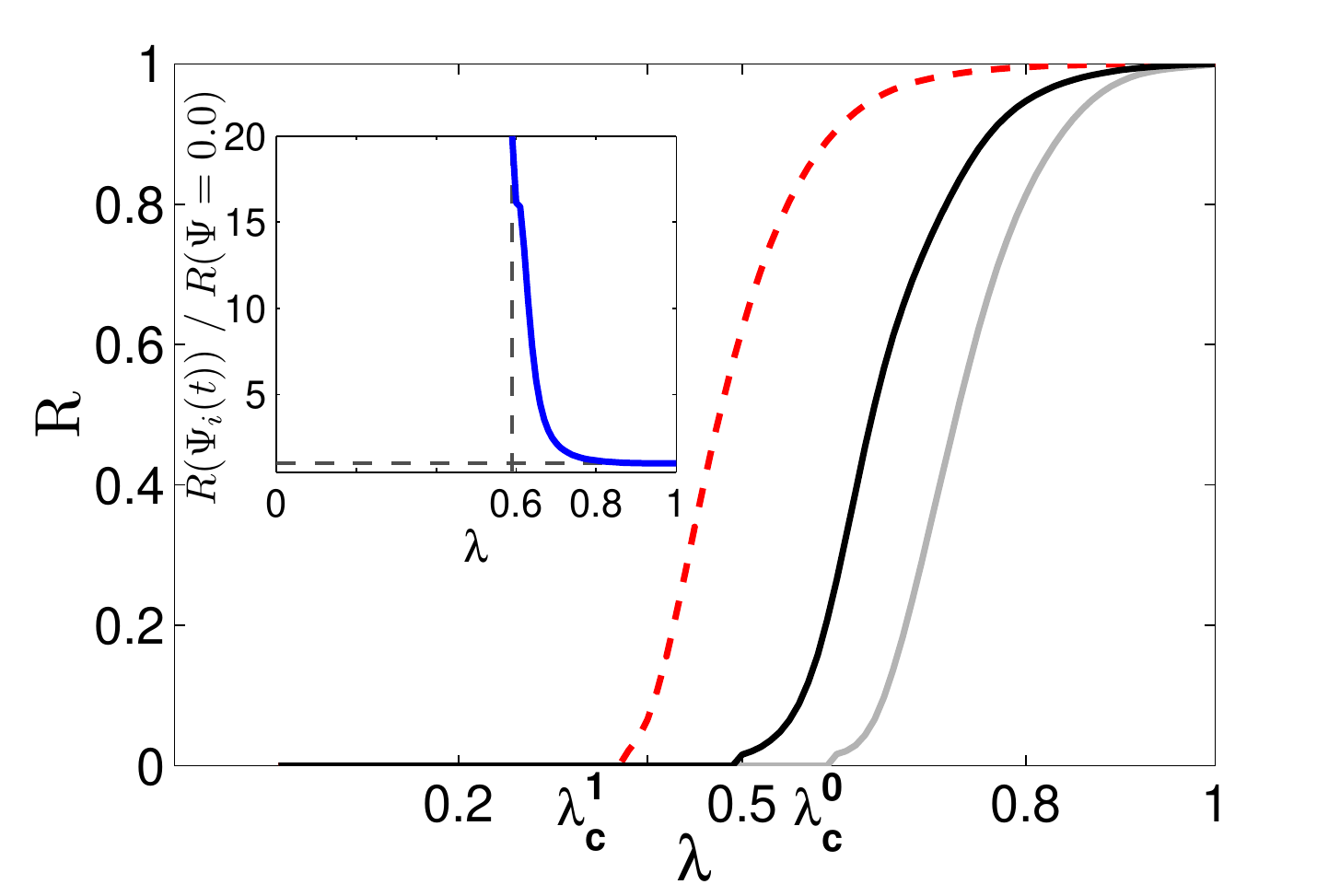}}   
  \subfloat[\emph{Facebook}]{\label{fig:FACE}\includegraphics[width= 0.3\textwidth]{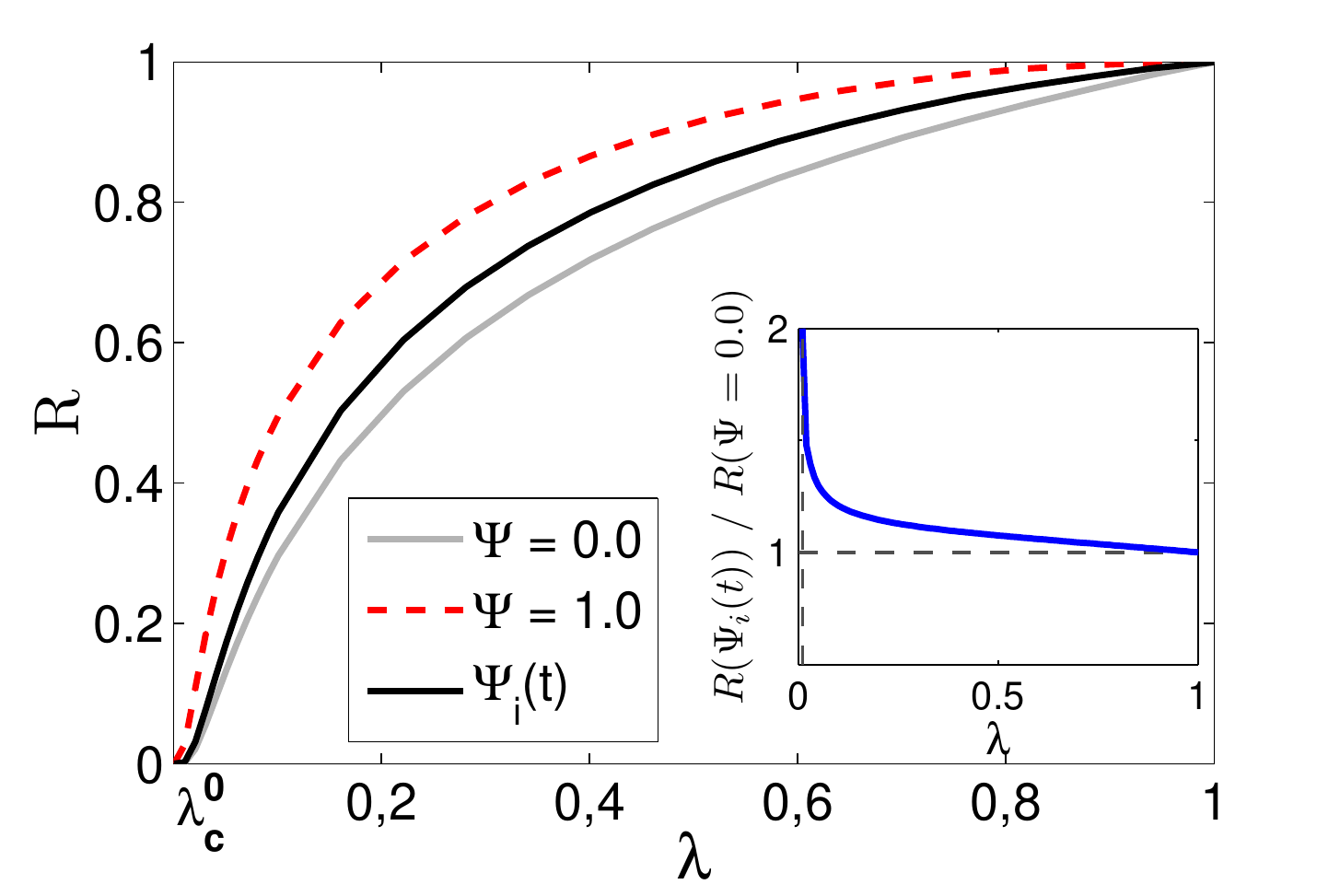}}   
\caption{\label{fig:lambdamodel} 
 (Color online) Phase diagram of the macroscopic density of stiflers as a function of parameter $\lambda$ for (a) a Barab{\'a}si-Albert (BA), (b) Erd{\"o}s-R{\'e}nyi (ER), (c) spatial scale-free (SSF), (d) \emph{Google+}, (e) \emph{USAroad} and (f) \emph{Facebook} networks. The artificial networks have the same size $N = 1000$ and average degree $ \langle k \rangle = 11.9 $, and the real-world networks are described in Table~\ref{tab:DescricaoDasBasesUtilizadas}. The classical propagation is recovered when $\Psi = 0$, whereas the complete curiosity case occurs for $\Psi= 1$. $\Psi_i(t)$ means the dynamical curiosity over time (Figure~\ref{fig:exemplo}). The $R$ values are the average of the final density of stiflers among all vertices. The inset figures are the quotient between the results of $R$ considering the dynamical curiosity and the respective $R$ of the classical propagation.}\vspace{-0.5cm}
\end{figure*}

The impact of the curiosity in the propagation process was analyzed on top of three network models, presenting different degree distributions, and three real-world networks. We consider the Barab{\'a}si-Albert~\cite{barabasiEalbert1999}  (BA), Erd{\"o}s-R{\'e}nyi~\cite{erdosrenyi1959}  (ER)  and Barth{\'e}lemy spatial scale free~\cite{Barthelemy2003} (SSF)  models, and for real-world networks we adopt the \emph{Google+}~\cite{GplusPaper}, \emph{Facebook}~\cite{viswanath09} and \emph{USAroad}~\cite{Barbieri2011}, the road network of USA. Their topological features are summarized in Table~\ref{tab:DescricaoDasBasesUtilizadas}, with the corresponding average degree $\left\langle k \right\rangle$, second moment of degree distribution $\left\langle k^2 \right\rangle$, average shortest path length $\left\langle \ell \right\rangle$ and assortativity coefficient that measures degree-degree correlation $\rho$. For the real-world networks we take the main component and assumed them as undirected networks. 


\subsection{Simulation Setup}

We analyze the proposed model of information spreading considering the macroscopic scale of the propagation. For this purpose, the initial setup for each simulation is $\dIg(0)  =  1 -  1/{N}$, $\dSp(0) =  {1}/{N} $ e $\dSt(0)  =  0$.  We calculate the corresponding density of stiflers $\dSt (t)$ and spreader $ \dSp(t)$ over time by taking each vertex as the only initial seed. For this value was considered an average of $60$ realizations.  The macroscopic density of stiflers $\dSt$ is calculated by averaging the number of stifler among all nodes. The macroscopic measures represent the propagation values for the whole network. Without lack of generality, we adopt the recovery probability $\mu = 1$~\cite{Pastor-Satorras2015,Vega-OliverosPRE2017}. 



\subsection{Simulation results}

\begin{figure*}[!t]
\centering	
  \subfloat[]{\label{fig:BA}\includegraphics[width= 0.3\textwidth]{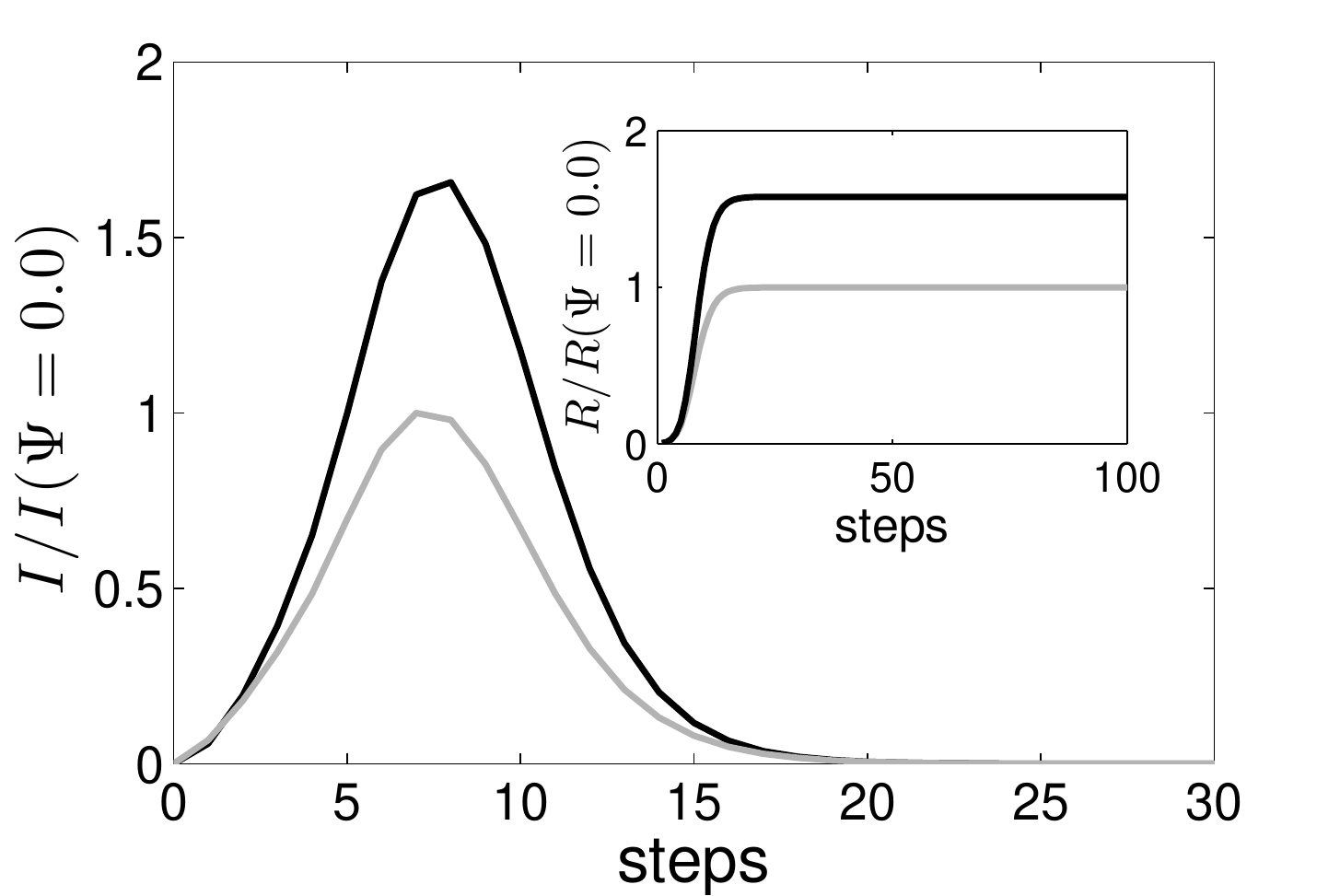}}   
  \subfloat[]{\label{fig:ER}\includegraphics[width= 0.3\textwidth]{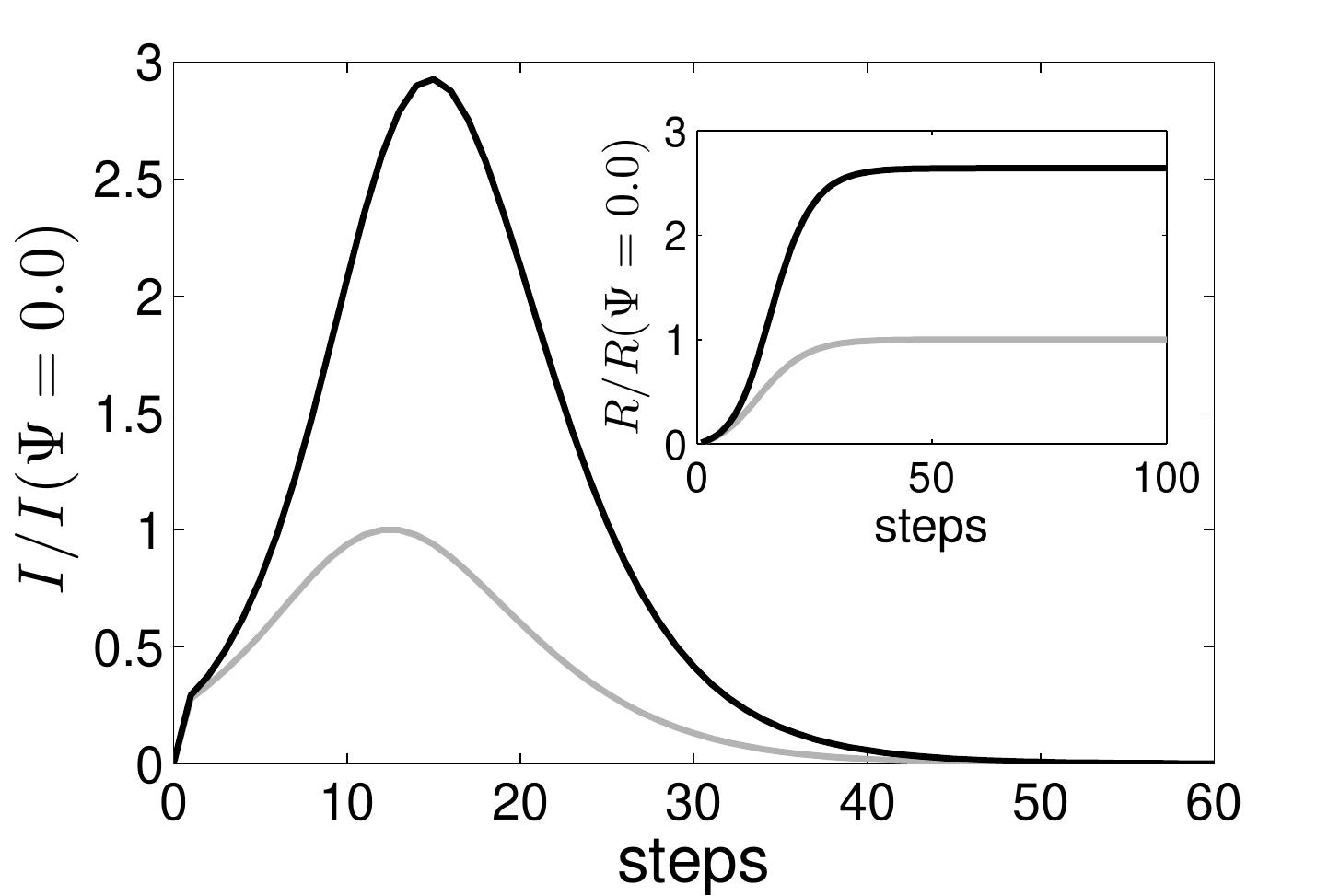}}  
  \subfloat[]{\label{fig:SSF}\includegraphics[width= 0.3\textwidth]{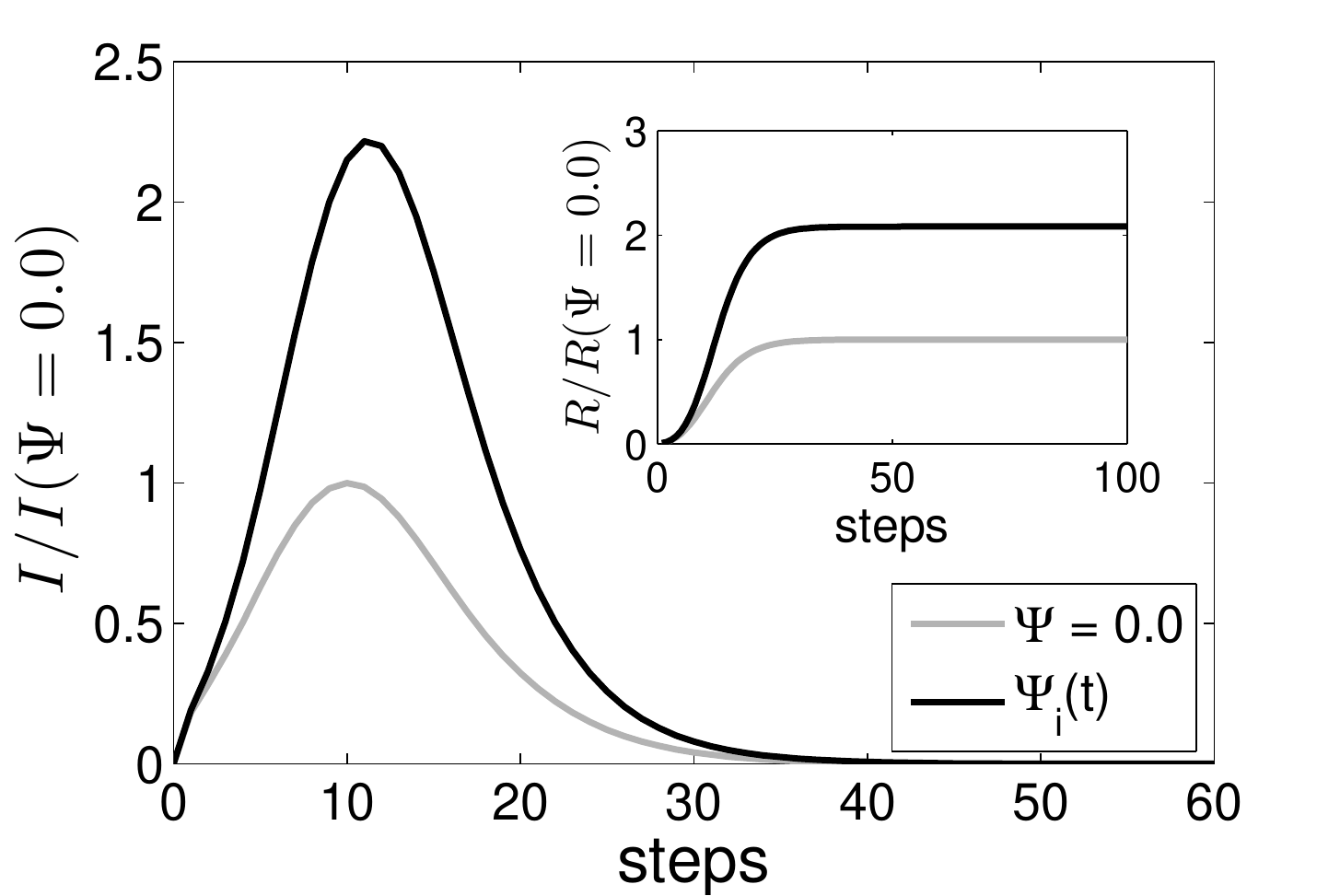}}\vspace{-0.3cm}   

\subfloat[]{\label{fig:gplus}\includegraphics[width= 0.3\textwidth]{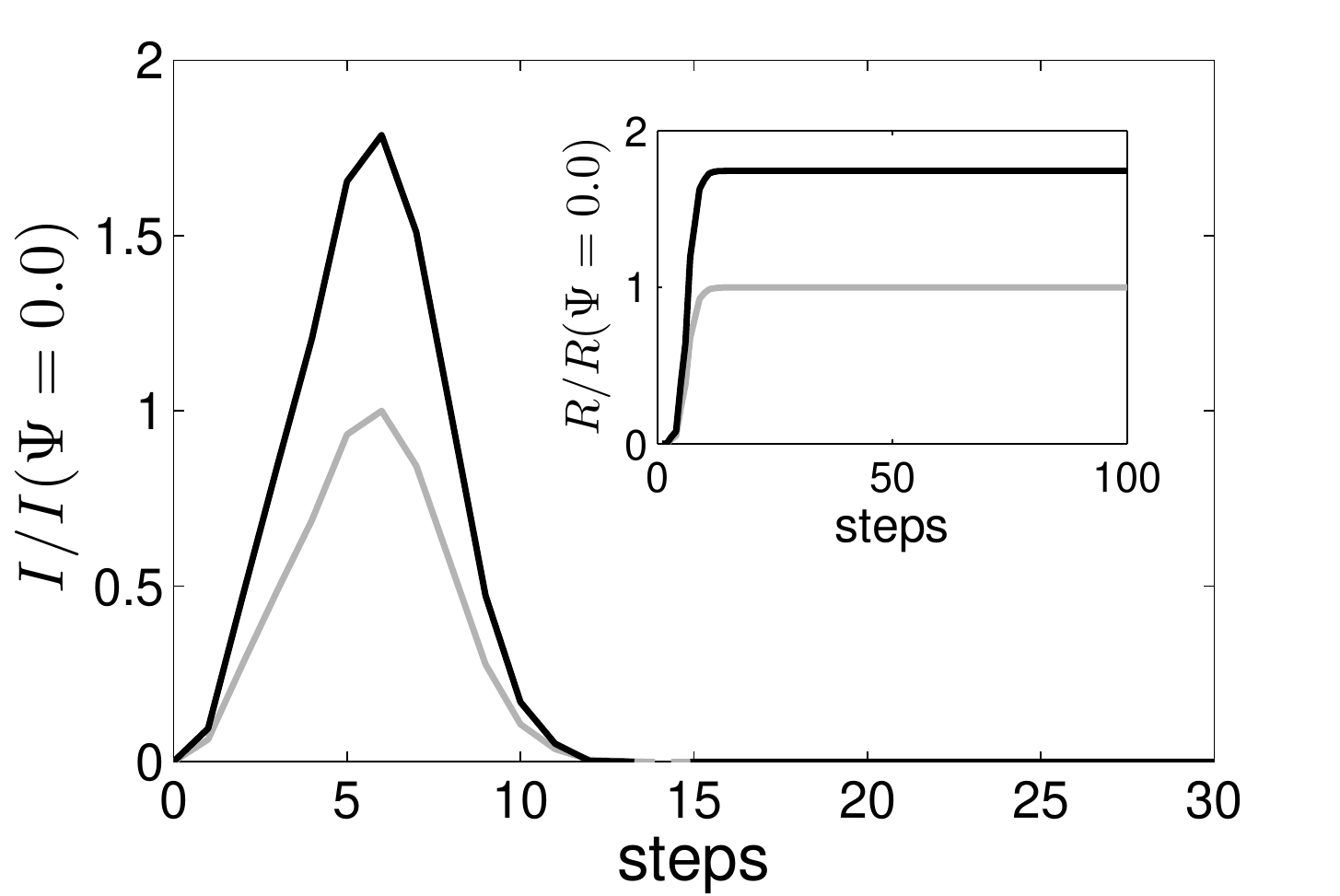}}   
  \subfloat[]{\label{fig:usa}\includegraphics[width= 0.3\textwidth]{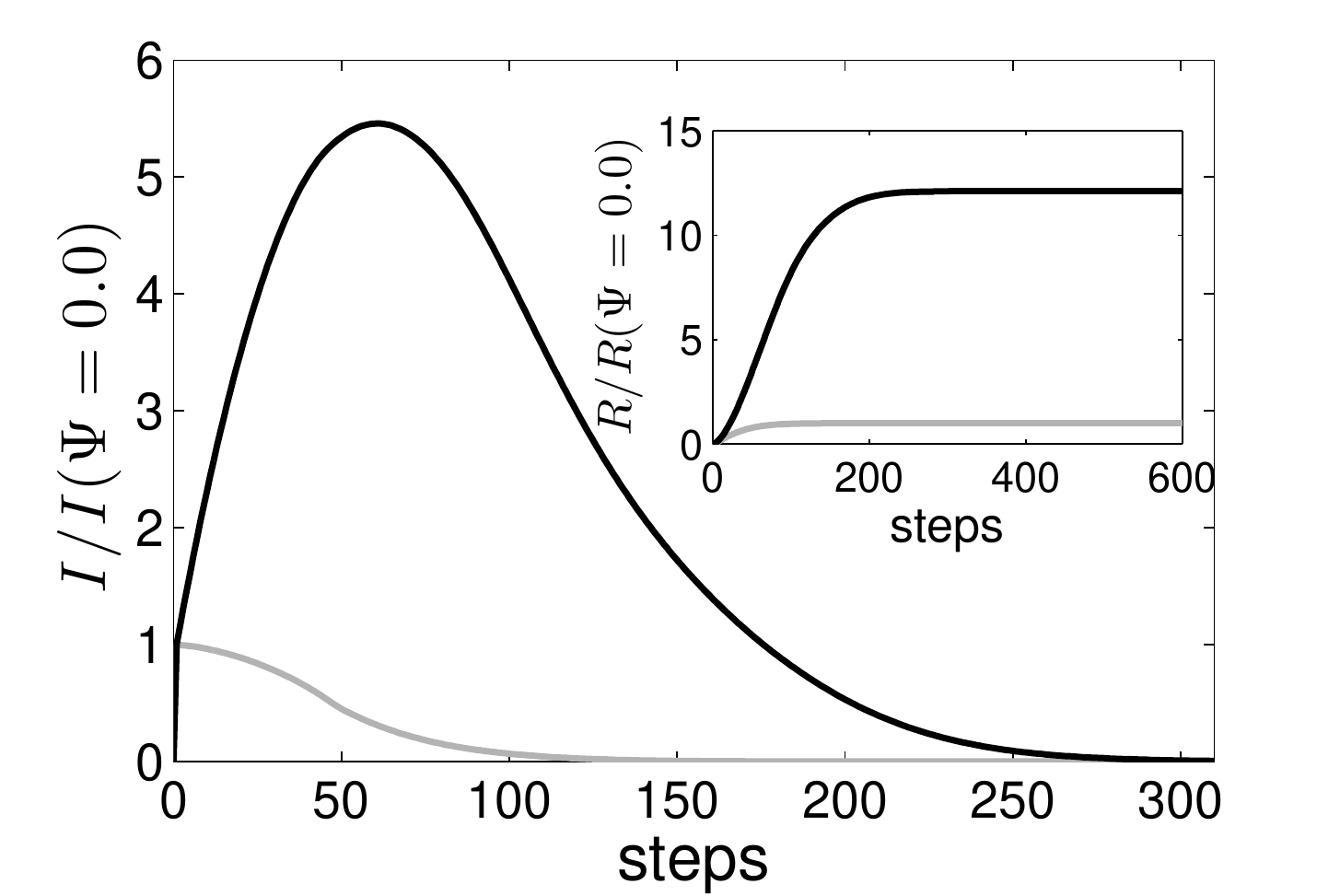}}  
  \subfloat[]{\label{fig:face}\includegraphics[width= 0.3\textwidth]{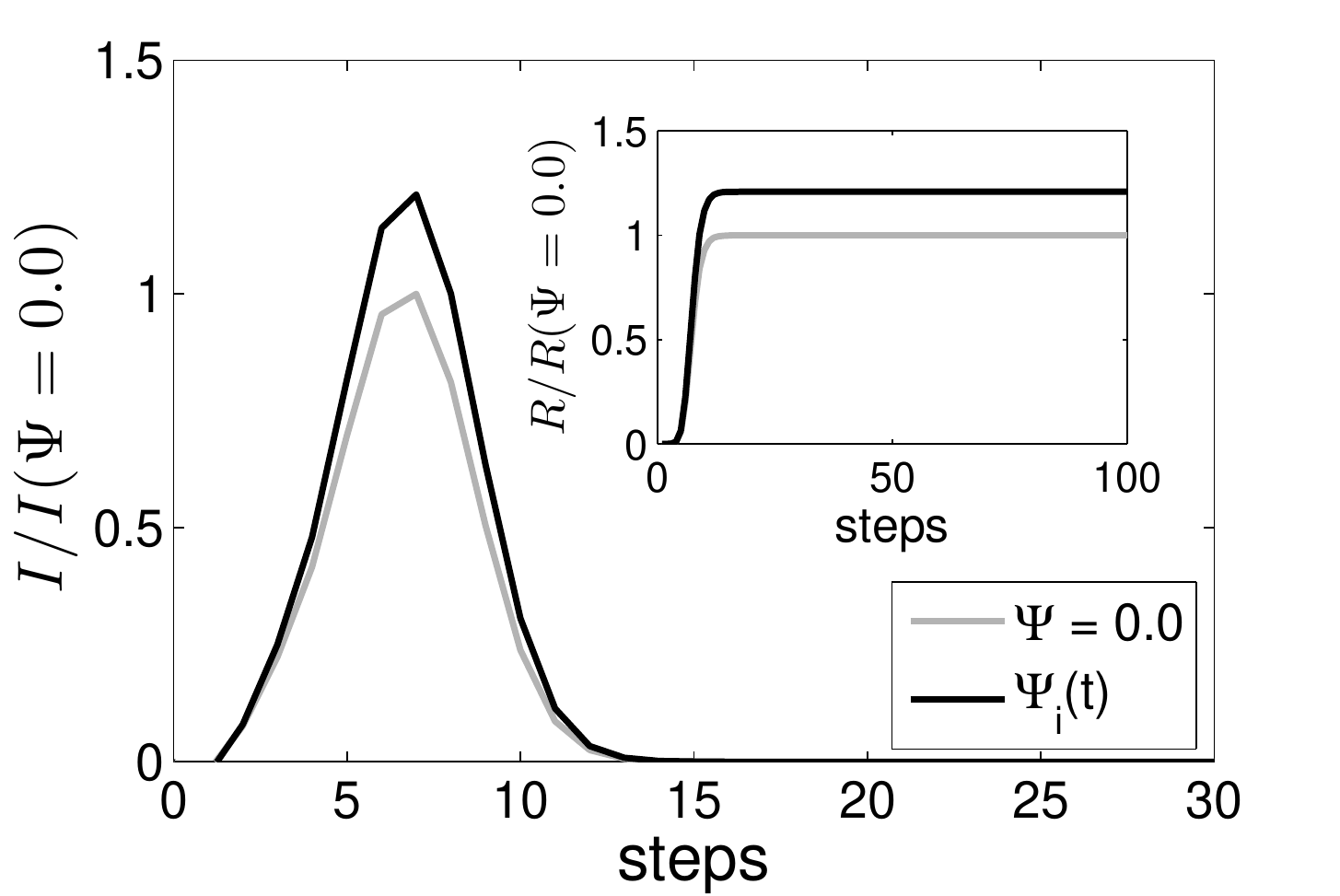}} 
\caption{\label{fig:fractionmodel} 
 (Color online) Macroscopic density of spreaders and stiflers over time  for (a) a Barab{\'a}si-Albert (BA), (b) Erd{\"o}s-R{\'e}nyi (ER), (c) spatial scale-free (SSF), (d) \emph{Google+}, (e) \emph{USAroad} and (f) \emph{Facebook} networks. Parameter $\lambda$ was fixed to a value $0.61$ for \emph{USAroad} and $0.1$ for the remaining networks. In the figures we have the curves of the average peak of spreaders and the insets are the corresponding average density of stiflers over time. The averages are calculated considering each vertex as a initial seed.}\vspace{-0.3cm}
\end{figure*}

The impact of the social curiosity is evaluated on the information spreading process. Initially, we analyze whether the phase transition for the endemic state is affected by the inclusion of the curiosity parameter. For different values of $0 < \lambda \leq 1$, the average density of stifler and the critical threshold are evaluated, as shown in Figure \ref{fig:lambdamodel}. 

We observe that independently of the network structure, i.e., scale-free, spatial or Poisson networks, the curiosity phenomenon improves the overall propagation. For the artificial networks (Fig. \ref{fig:modelBA}, \ref{fig:modelER}, \ref{fig:modelSSF}), the critical threshold of the Utopian case  ($\lambda_c^{1}$) is very close to half of the corresponding threshold of the non-curiosity configuration ($\lambda_c^{0}$), when $\Psi = 0$. In the real-world networks, the critical thresholds $\lambda_c^{0}$ and $\lambda_c^{1}$ are very similar and tend to zero, (Fig.~\ref{fig:GPLUS}, \ref{fig:FACE}). This occurs due to real-world networks have a scale-free organization, characterized by a power-law degree distribution~\cite{barabasiEalbert1999,newman2010networks} ($P(k) \sim k^{-\gamma}$ with $\gamma < 3$). In these networks, the critical threshold $\lambda_c \rightarrow 0$ when $N \rightarrow \infty$. The \emph{USAroad} is the most homogeneously distributed network, having low variance in the degree distribution and larger average shortest path length (Table~\ref{tab:DescricaoDasBasesUtilizadas}). 
The dynamical curiosity ($\Psi_i(t)$) little affects the critical threshold in comparison with the non-curiosity case ($\Psi = 0$). The final density of stiflers is improved in more than twice than the results reached by the non-curiosity case in lower values of $\lambda$, and tend to be ineffective for larger values (inset Figure~\ref{fig:lambdamodel}). 

\begin{figure}[!ptb]
\centering	
\subfloat[]{\label{fig:artLambdas}\includegraphics[width= 0.35\textwidth]{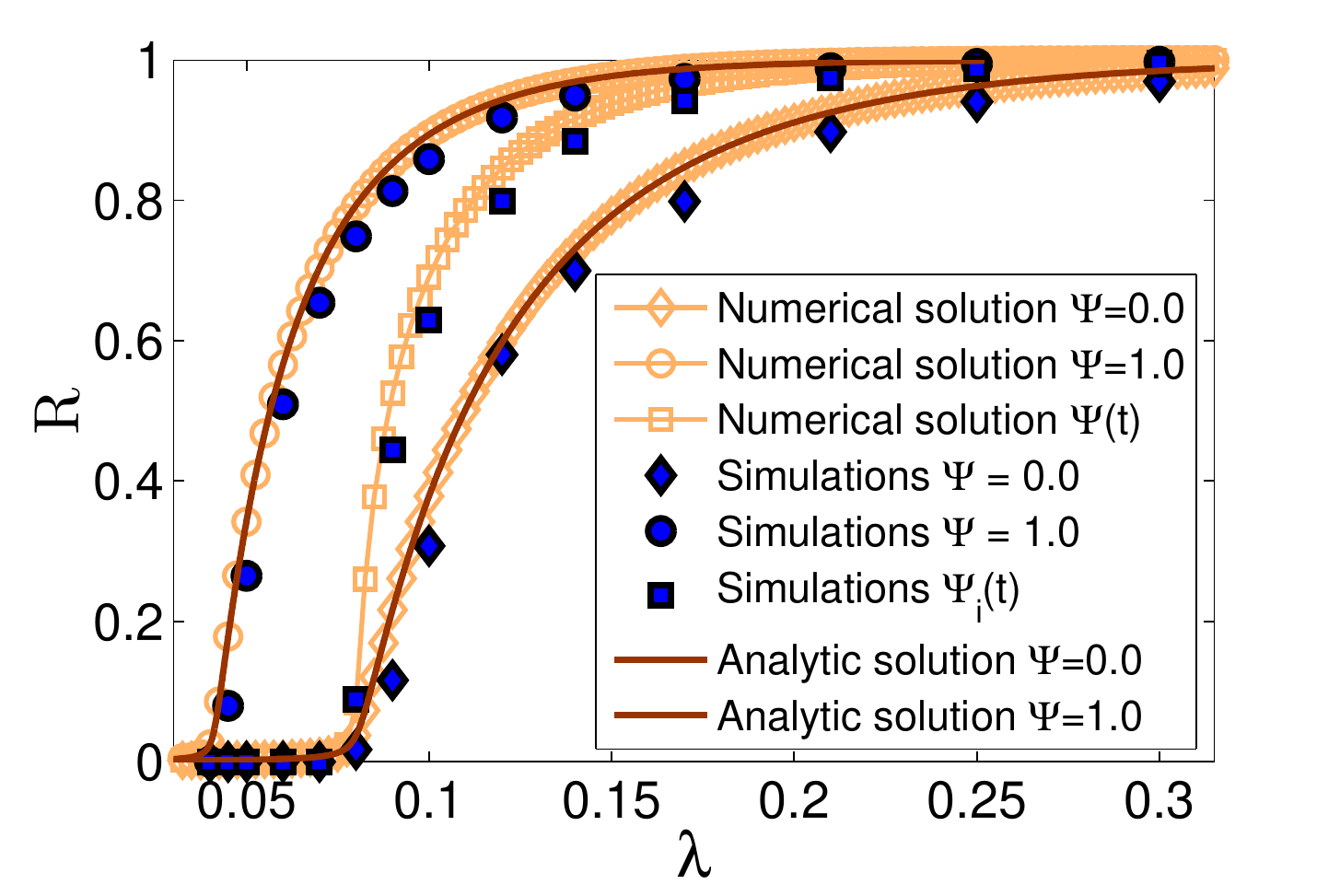}} \vspace{-0.3cm}  

\subfloat[]{\label{fig:artDif}\includegraphics[width= 0.35\textwidth]{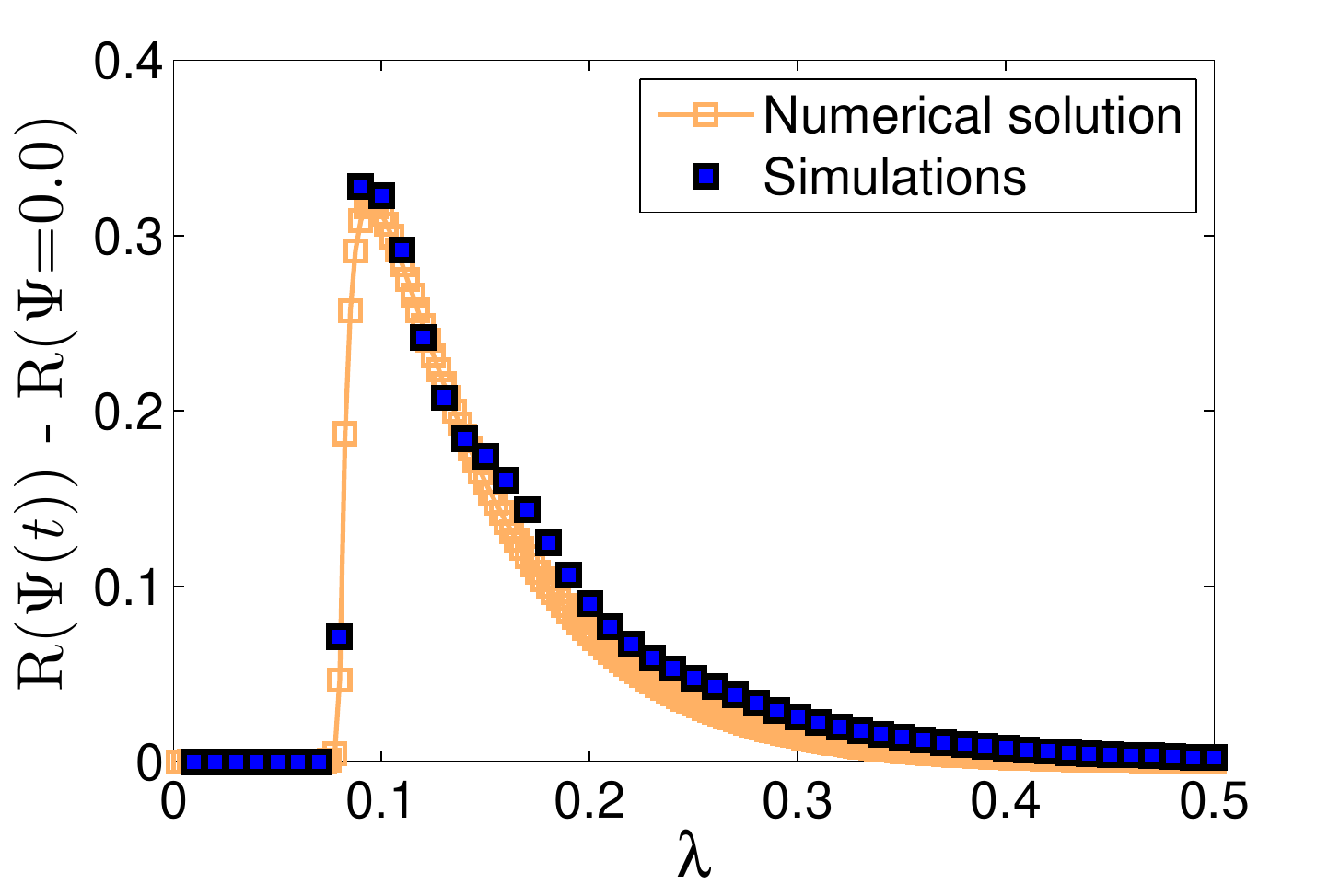}}   
\caption{\label{fig:lambdaTheory} 
 (Color online)  (a) Final density of stiflers $R$ vs $\lambda=\beta/\mu$ for curiosity $\Psi=0.0$ (diamonds), $\Psi=1.0$ (circles) and $\Psi(t)=I(t)+R(t)$ (squares).  Filled symbols correspond to Monte Carlo simulations on Erd\"os-Renyi networks of size $N=10^4$ with initial density of spreaders $I(0)=10^{-3}$, whereas empty symbols represent the numerical integration of Eqs.~(\ref{SIR-t}).  Solid lines are the analytical solutions from Eqs.~(\ref{beta-R-psi}).  (b) Difference between the final density of stiflers in the dynamic case $\Psi(t)$ and in the $\Psi=0$ case, obtained from the data of panel (a).}\vspace{-0.3cm} 
\end{figure}

To study the evolution of the diffusion process, we evaluate the curves of spreaders achieved in each network, and the insets are the corresponding density of stiflers (Figure~\ref{fig:fractionmodel}). Parameter $\lambda$ is fixed to a value close to $\lambda_c^{0}$, which is $0.61$ for \emph{USAroad} and $0.1$ for the remaining networks. The peak of spreaders and the final density of stiflers for the social curiosity are superior in all the case to the classical spreading approach ($\Psi = 0$). The $\Psi_i(t)$ curves of the peak of spreaders and density of stiflers last the same time than the classical approach. This indicates that the curiosity can improve the extent and velocity of the propagation. Specifically, the peak of the spreaders is twice or higher the respective peaks of the classical approach, except for the \emph{Facebook} network, which is $23\%$ higher. The $\Psi_i(t)$ curves also achieve more than twice the final density of stiflers of the $\Psi = 0.0$ curves, for the \emph{ER}, \emph{SSF} and \emph{USAroad} networks (inset Fig.~\ref{fig:ER}, \ref{fig:SSF},\ref{fig:usa}); it achieves $50\%$ or higher density of stiflers in the \emph{BA} and \emph{Google+} networks (inset Fig.~\ref{fig:BA}, \ref{fig:gplus}); and it is $20\%$ higher in the \emph{Facebook} network (inset Fig~\ref{fig:face}). 

The results suggest that adopting the social curiosity in the diffusion process can improve the expected density of informed individuals, as well, it increases the peak of spreaders and the velocity of the dynamic. The impact in the results depend on the network topology and the effective spreading rate $\lambda$ employed. For instances, for spatially distributed networks like \emph{USAroad}, we obtain very significant results than in social networks like \emph{Google+} or \emph{Facebook}.

\section{Mean-field analysis}
\label{mean-field}

The system behavior can be described, at the mean-field level, by the following set of coupled recurrence equations for the time evolution of the densities of individuals $\dIg$, $\dSp$ and  $\dSt$: 
\begin{subequations}
\begin{alignat}{3}
\label{St}
S(t+1)&=S(t)-\gamma S(t),  \\
\label{It}
I(t+1)&=I(t)+\gamma S(t) - \mu I(t), \\
\label{Rt}
R(t+1)&=R(t)+ \mu I(t), 
\end{alignat}{}
\label{SIR-t}
\end{subequations} which allows calculating their values at time $t+1$ from the previous time $t$.  These equations describe an infinitely large system, where fluctuations due finite-size effects are neglected.  Here 
\begin{equation}
\gamma(t) \equiv 1-[(1-\beta)(1- \beta \Psi(t))]^{\langle k \rangle I(t)}
\label{gamma}
\end{equation}
is the probability that a susceptible vertex gets informed from one of its spreader neighbors, and $\Psi(t)$ is the curiosity level, i. e., the probability that a susceptible vertex requests the information to each spreader.  Within a mean-field approach, the curiosity is approximated as the mean fraction of informed neighbors of a given vertex $i$, $\Psi(t) \simeq I(t)+R(t)$, while the exponent $\langle k \rangle I(t)$ in $\gamma$ is the estimated mean number of spreader neighbors of $i$.  In Eq.~(\ref{It}), the gain term $\gamma S(t)$ represents the fraction of susceptible nodes that become spreaders, while the loss term $-\mu I(t)$ corresponds to the fraction of spreaders that become stiflers in a time step.  

In Fig.~\ref{fig:lambdaTheory}(a) we plot the stationary value of $R$ vs $\lambda=\beta/\mu$ obtained from Monte Carlo simulations (filled symbols) and compare with results from the numerical integration of Eqs.~(\ref{SIR-t}) (empty symbols).  Simulations were performed on Erd\"os-Renyi networks of $N=10^4$ nodes, starting with a small fraction $I(0)=10^{-3}$ of spreaders uniformly distributed over the network.  Diamonds, squares and circles correspond to curiosity $\Psi=0$, $\Psi(t)=I(t)+R(t)$ and $\Psi=1$, respectively.  We observe that the agreement between theory and simulations is quite good for the entire range of $\lambda$.  The mean-field solution is also able to capture the transition point $\lambda_c$ at which the final density of stiflers becomes larger than zero, and gives the same qualitative behavior as the corresponding curves for all networks shown in Fig.~\ref{fig:lambdamodel}.  Figure~\ref{fig:lambdaTheory}(b) shows the difference in the values of $R$ corresponding to the dynamic case $\Psi(t)$ and $\Psi=0$.  We observe that $R(\Psi(t))$ is larger than $R(\Psi=0)$ for the entire range of $\lambda$, and that the difference becomes very large close to the transition point.  This shows that the spreading of the information is largely increased when the curiosity mechanism is taken into account. 

We now perform the analysis of Eqs.~(\ref{SIR-t}) starting from the simplest case in which the curiosity $\Psi$ is constant over time.  To solve Eqs.~(\ref{SIR-t}) we make two approximations.  First, we approximate the difference $S(t+1)-S(t)$ by the time derivative $\frac{dS(t)}{dt}$ in the continuous time limit, and analogously for $I$ and $R$.  Next, we assume that $I(t)$ is small and expand $\gamma(t)$ from Eq.~(\ref{gamma}) to first order in $I \ll 1$ and obtain 
\begin{equation}
\gamma(t) \simeq - \langle k \rangle \ln[(1-\beta)(1-\Psi \beta)] I(t) \label{gamma-1}.
\end{equation}
Then, Eqs.~(\ref{SIR-t}) become
\begin{subequations}
\begin{alignat}{3}
\label{dSdt}
\frac{dS(t)}{dt}&= \langle k \rangle \ln[(1-\beta)(1-\Psi \beta)] I(t) S(t) \\
\label{dIdt}
\frac{dI(t)}{dt}&= -\langle k \rangle \ln[(1-\beta)(1-\Psi \beta)] I(t) S(t)- \mu I(t), \\
\label{dRdt}
\frac{dR(t)}{dt}&= \mu I(t).
\end{alignat}{}
\label{SIR-t-1}
\end{subequations}
Dividing Eq.~(\ref{dSdt}) by Eq.~(\ref{dRdt}) we arrive to the equation
\begin{equation}
\frac{dS}{dR} = \frac{\langle k \rangle}{\mu} \ln[(1-\beta)(1-\Psi \beta)] S,
\label{dSdR}
\end{equation}
which gives a relation between $S$ and $R$. Then, the integration of Eq.~(\ref{dSdR}) gives, after some algebra, the following equation that relates $S$ and $R$ at all times:
\begin{equation}
(1-\beta)(1-\Psi \beta) =  
\left( \frac{S}{S_0} \right)^{\frac{\mu}{\langle k \rangle R}}
\end{equation}
where we have used the initial condition $S_0=S(t=0)=1-1/N$ and $R(t=0)=0$.  At the stationary state ($t = \infty$), $R$ and $S$ take the values $R(t=\infty)=\tilde{R}$ and $S(t=\infty)=\tilde S=1-\tilde R$, respectively, since $I(t=\infty)=0$.  Then, the final density of stiflers obeys the equation 
\begin{equation}
(1-\beta)(1-\Psi \beta) =  
\left( \frac{1-R_{\infty}}{S_0} \right)^{\frac{\mu}{\langle k \rangle \tilde R}}.
\end{equation}
Finally, solving for $\beta$ leads to the expression 
\begin{equation}
\beta=\frac{(1+\Psi)-\sqrt{(1+\Psi)^2-4 \Psi 
\left[1-\left( \frac{1-\tilde R}{S_0} \right)^{\frac{\mu}{\langle k \rangle \tilde R}} \right]}}{2 \Psi},
\label{beta-R}
\end{equation}
which relates the stationary value of $R$ with the propagation force $\beta$.  For the extreme cases $\Psi=0$ and $\Psi=1$, Eq.~(\ref{beta-R}) is reduced to
\begin{eqnarray}
\beta&=& 1-\left( \frac{1-\tilde R}{S_0} \right)^{\frac{\mu}{\langle k \rangle \tilde R}} ~~~~ \mbox{for $\Psi=0$ ~~~ and} \nonumber \\
\beta&=& 1-\left( \frac{1-\tilde R}{S_0} \right)^{\frac{\mu}{2\langle k \rangle \tilde R}} ~~~~ \mbox{for $\Psi=1$}. 
\label{beta-R-psi}
\end{eqnarray}
In Fig.~\ref{fig:lambdaTheory}(a) we plot by solid lines the analytical solution $\tilde R$ vs $\lambda=\beta/\mu$ from Eqs.~(\ref{beta-R-psi}).  We observe a very good agreement with the numerical integration of Eqs.~(\ref{SIR-t}) (empty symbols).  An interesting relation between the two curves can be found from Eqs.~(\ref{beta-R-psi}), that is, $\tilde R(\Psi=1,\beta)=\tilde R(\Psi=0,1-(1-\beta)^2)$.  This means that the shape of $\tilde R$ vs $\lambda$ for $\Psi=1$ is the same as that for $\Psi=0$ with the x-axis shifted to the right by the factor $[1-(1-\mu \lambda)^2]/\mu$.

Another quantity of interest is the transition point $\beta_c$ between an ``endemic phase" where the information spreads to a large fraction of the population for $\beta>\beta_c$, and a ``healthy phase" where the information does not propagate for $\beta<\beta_c$.  Starting from a situation with a very small fraction of spreaders $I \ll 1$ and no stiflers $R=0$, Eq.~(\ref{dIdt}) becomes 
\begin{equation}
\frac{dI(t)}{dt} = \delta I(t), 
\label{dIdt-delta}
\end{equation}
with $\delta \equiv -\langle k \rangle \ln[(1-\beta)(1-\Psi \beta)]-\mu$, and where we have neglected terms of order $I^2$.  For $\delta>0$ ($\delta<0$), $I(t)$ grows (decays) exponentially fast.  Then, at the transition point is 
$\delta=-\langle k \rangle \ln[(1-\beta_c)(1-\Psi \beta_c)]-\mu=0$, from where arrive to the following expression for $\lambda_c=\beta_c/\mu$:
\begin{equation}
\lambda_c^{\Psi} = \frac{1+\Psi-\sqrt{(1+\Psi)^2-4 \Psi \left(1-e^{-\mu/\langle k \rangle}\right)}}{2 \Psi \mu}.
\end{equation}
For $\Psi=0$ and $\Psi=1$ we obtain, respectively, \\ $\lambda_c^0=[1-e^{-\mu/\langle k \rangle}]/\mu$ and  
$\lambda_c^1=[1-\sqrt{e^{-\mu/\langle k \rangle}}]/\mu$.  We can easily check that $\lambda_c^0=[1-(1-\mu \lambda_c^1)^2]/\mu$, in agreement with the shift of the $\tilde R$ vs $\lambda$ curves mentioned above. 

We now study the case where the curiosity changes over time according to $\Psi(t)=I(t)+R(t)$.  Expanding $\gamma$ from Eq.~(\ref{gamma}) to first order in $I$ we obtain \\ 
$\gamma(t) \simeq - \langle k \rangle \ln[(1-\beta)(1-\beta R(t))] I(t)$.  Then, in the continuous time limit Eqs.~(\ref{SIR-t}) become 
\begin{subequations}
\begin{alignat}{3}
\label{dSdt-2}
\frac{dS(t)}{dt}&= \langle k \rangle \ln[(1-\beta)(1-\beta R(t))] I(t) S(t) \\
\label{dIdt-2}
\frac{dI(t)}{dt}&= -\langle k \rangle \ln[(1-\beta)(1-\beta R(t))] I(t) S(t)- \mu I(t), \\
\label{dRdt-2}
\frac{dR(t)}{dt}&= \mu I(t).
\end{alignat}{}
\label{SIR-t-2}
\end{subequations}
Following an approach similar to the one above for constant $\Psi$, the transition point is obtained by linearizing Eq.~(\ref{dIdt-2}) around the fixed point $I=0$, $R=0$.  This leads to an equation like Eq.~(\ref{dIdt-delta}) with the same value of $\delta$.  Therefore, the transition point with dynamical curiosity $\Psi(t)$ is the same as that with $\Psi=0$.  This can be observed in Fig.~\ref{fig:lambdaTheory}(a) and also in Fig.~\ref{fig:lambdamodel} for all networks, except for the USA road network [panel (e)].

\section{Discussion and Final Remarks} \label {conclusions}

The curiosity is an intrinsic feature that varies from person to person.  The level of curiosity of people can be increased not only among the most curious subjects but also among the less curious individuals that are willing to adopt a new product or idea.  More marketing campaigns are employing curiosity strategies to attract customers~\cite{websiteCU,websitePG}. For instance, the game Pok{\'{e}}mon GO, the biggest mobile game in US history~\cite{websitePG}, introduces virtual reality and outdoor activities in public locations, which increase people's curiosity. Another example is Philips R{\'{e}}Aura, a skin rejuvenation brand where the customers can access an interactive site that shows diaries and trial results from real people~\cite{websiteCU}. The social platform for customer engagement and interaction promotes users' curiosity about the truth behind the product claims.  
Stimulation of people's curiosity can provide better results not only for the marketing area but also for education and political or health campaigns, among others. In general, people's opinions are influenced by their close contacts, i. e., family and friends, and that are why they tend to spread these ideas more than those they receive from other sources outside of the circle of close relationships.   

This work contributes to understanding how social curiosity can improve the potential diffusion of information on networks.  We proposed an information spreading model based on the $\SIR$ dynamics and introduced the social curiosity as a heterogeneous dynamical parameter that evolves over time. Depending on the fraction of informed neighbors, susceptible individuals become curious and interact with the neighborhood to satisfy the curiosity about the new idea or information. 
We found that the effect of the social curiosity on the propagation of the information depends on the effective spreading rate $\lambda$.  For values of $\lambda$ above and close to the healthy-endemic transition point $\lambda_c$, the curiosity is able to increase the propagation level in more than twice compared to the classical approach without curiosity.  The maximum number of new spreaders reached during the propagation process and the speed of propagation are also increased by the curiosity.  However, as $\lambda$ becomes smaller or larger than $\lambda_c$,  the social curiosity has no significant effect.  These results suggest that the curiosity mechanism reaches its maximum effectiveness  when the system is just above the transition point, in the endemic phase.  Another important point is the structure of the network.  We found that the curiosity has a larger impact on homogeneous and spatial networks than on on-line social and scale-free networks.

The proposed curiosity mechanism might be applicable to other fields like biological networks and animal social behavior models.  Another important issue to be studied is the interplay between the curiosity influence --like location-base strategies, the impact of network structure and dynamical parameters, and the personality homophily  for the adoption/diffusion of ideas and publicity.  More suitable models can be developed adopting the social curiosity and analyzing geolocated phenomena like the Pok{\'{e}}mon GO game. 


\section*{Acknowledgments}

{\small This research employed the computing resources of the Center for Mathematical Sciences Applied to Industry (CeMEAI), financed by FAPESP. DAVO acknowledges CNPq (grant 140688/2013-7), FAPESP (grant 2016/23698-1) and the Banco Santander grant of RED Macro Universidades. LB acknowledges CAPES and FAPESP. FV acknowledges support from CONICET (Argentina).  FAR acknowledges CNPq (grant 305940/2010-4) and FAPESP (grant 13/26416-9).
}


\bibliographystyle{IEEEtran}
\balance

\bibliography{bibliography}

\begin{thebibliography}{10}
\providecommand{\url}[1]{#1}
\csname url@samestyle\endcsname
\providecommand{\newblock}{\relax}
\providecommand{\bibinfo}[2]{#2}
\providecommand{\BIBentrySTDinterwordspacing}{\spaceskip=0pt\relax}
\providecommand{\BIBentryALTinterwordstretchfactor}{4}
\providecommand{\BIBentryALTinterwordspacing}{\spaceskip=\fontdimen2\font plus
\BIBentryALTinterwordstretchfactor\fontdimen3\font minus
  \fontdimen4\font\relax}
\providecommand{\BIBforeignlanguage}[2]{{%
\expandafter\ifx\csname l@#1\endcsname\relax
\typeout{** WARNING: IEEEtran.bst: No hyphenation pattern has been}%
\typeout{** loaded for the language `#1'. Using the pattern for}%
\typeout{** the default language instead.}%
\else
\language=\csname l@#1\endcsname
\fi
#2}}
\providecommand{\BIBdecl}{\relax}
\BIBdecl

\bibitem{Noe2016}
N.~No{\"{e}}, R.~M. Whitaker, and S.~M. Allen, ``Personality homophily and the
  local network characteristics of facebook,'' in \emph{2016 IEEE/ACM
  International Conference on Advances in Social Networks Analysis and Mining
  (ASONAM)}, Aug 2016, pp. 386--393.

\bibitem{Tabacchi2017}
M.~E. Tabacchi, B.~Caci, M.~Cardaci, and V.~Perticone, ``Early usage of
  pok{\'{e}}mon go and its personality correlates,'' \emph{Computers in Human
  Behavior}, vol.~72, pp. 163 -- 169, 2017.

\bibitem{barabasiEalbert1999}
A.-L. Barab{\'{a}}si and R.~Albert, ``{Emergence of scaling in random
  networks},'' \emph{Science}, vol. 286, pp. 509--512, 1999.

\bibitem{newman2010networks}
M.~Newman, \emph{{Networks: an introduction}}.\hskip 1em plus 0.5em minus
  0.4em\relax Oxford Uni. Press, Inc., 2010.

\bibitem{Guille2013}
A.~Guille, H.~Hacid, C.~Favre, and D.~A. Zighed, ``{Information diffusion in
  online social networks: A survey},'' \emph{ACM SIGMOD Record}, vol.~42,
  no.~1, p.~17, 5 2013.

\bibitem{Pastor-Satorras2015}
R.~Pastor-Satorras, C.~Castellano, P.~Van~Mieghem, and A.~Vespignani,
  ``{Epidemic processes in complex networks},'' \emph{Reviews of Modern
  Physics}, vol.~87, no.~3, pp. 925--979, 8 2015.

\bibitem{Kempe15}
D.~Kempe, J.~Kleinberg, and v.~Tardos, ``{Maximizing the Spread of Influence
  through a Social Network},'' \emph{Theory of Computing}, vol.~11, no.~4, pp.
  105 -- 147, 2015.

\bibitem{porter2016}
M.~A. Porter and J.~P. Gleeson, \emph{{Dynamical Systems on Networks: A
  Tutorial}}, 1st~ed., ser. Frontiers in Applied Dynamical Systems: Reviews and
  Tutorials.\hskip 1em plus 0.5em minus 0.4em\relax Springer, 2016.

\bibitem{Vega-OliverosH17}
D.~A. Vega-Oliveros, L.~da~F~Costa, and F.~A. Rodrigues, ``Rumor propagation
  with heterogeneous transmission in social networks,'' \emph{Journal of
  Statistical Mechanics: Theory and Experiment}, vol. 2017, no.~2, p. 023401,
  2017.

\bibitem{ZUCKERMAN:1986}
M.~Zuckerman and P.~Litle, ``Personality and curiosity about morbid and sexual
  events,'' \emph{Personality and Individual Differences}, vol.~7, no.~1, pp.
  49 -- 56, 1986.

\bibitem{Litman:2005}
J.~A. Litman, ``Curiosity and the pleasures of learning: Wanting and liking new
  information,'' \emph{Cognition and Emotion}, vol.~19, no.~6, pp. 793--814,
  2005.

\bibitem{Wu:2013}
Q.~Wu and C.~Miao, ``Curiosity: From psychology to computation,'' \emph{ACM
  Comput. Surv.}, vol.~46, no.~2, pp. 18:1--18:26, 2013.

\bibitem{NYT:2016}
\BIBentryALTinterwordspacing
M.~I. Nick~Wingfield, ``Pokemon go brings augmented reality to a mass
  audience,'' July, 2016. [Online]. Available: \url{http://www.nytimes.com}
\BIBentrySTDinterwordspacing

\bibitem{Zhang2016}
H.~Zhang, A.~Kuhnle, H.~Zhang, and M.~T. Thai, ``Detecting misinformation in
  online social networks before it is too late,'' in \emph{2016 IEEE/ACM
  International Conference on Advances in Social Networks Analysis and Mining
  (ASONAM)}, Aug 2016, pp. 541--548.

\bibitem{Zhao2011}
L.~Zhao, Q.~Wang, J.~Cheng, Y.~Chen, J.~Wang, and W.~Huang, ``{Rumor spreading
  model with consideration of forgetting mechanism: A case of online blogging
  LiveJournal},'' \emph{Physica A: Statistical Mechanics and its Applications},
  vol. 390, no.~13, pp. 2619--2625, 2011.

\bibitem{Nekovee2007}
M.~Nekovee, Y.~Moreno, G.~Bianconi, and M.~Marsili, ``{Theory of rumour
  spreading in complex social networks},'' \emph{Physica A: Statistical
  Mechanics and its Applications}, vol. 374, no.~1, pp. 457--470, 1 2007.

\bibitem{Vega-Oliveros-socinf15}
D.~Vega{-}Oliveros, L.~Berton, A.~Lopes, and F.~Rodrigues, ``Influence
  maximization based on the least influential spreaders,'' in \emph{SocInf
  2015, co-located with {IJCAI} 2015}, vol. 1398, 2015, pp. 3--8.

\bibitem{Hethcote2006}
H.~W. Hethcote, ``\BIBforeignlanguage{en}{{The Mathematics of Infectious
  Diseases}},'' \emph{\BIBforeignlanguage{en}{SlAM REVIEW}}, 8 2006.

\bibitem{Buono2013}
C.~Buono, F.~Vazquez, P.~A. Macri, and L.~A. Braunstein, ``{Slow epidemic
  extinction in populations with heterogeneous infection rates},''
  \emph{Physical Review E}, vol.~88, no.~2, p. 022813, 2013.

\bibitem{Vega-OliverosPRE2017}
\BIBentryALTinterwordspacing
D.~A. Vega-Oliveros, L.~d.~F. Costa, and F.~A. Rodrigues, ``{Influence
  maximization on correlated networks through community identification},'' may
  2017. [Online]. Available: \url{http://arxiv.org/abs/1705.00630}
\BIBentrySTDinterwordspacing

\bibitem{viswanath09}
B.~Viswanath, A.~Mislove, M.~Cha, and K.~P. Gummadi, ``On the evolution of user
  interaction in {Facebook},'' in \emph{Proc. Workshop on Online Social
  Networks}, 2009, pp. 37--42.

\bibitem{Vega-OliverosB15}
D.~Vega{-}Oliveros and L.~Berton, ``Spreader selection by community to maximize
  information diffusion in social networks,'' in \emph{SIMBig 2015}, 2015, pp.
  73--82.

\bibitem{erdosrenyi1959}
A.~Erdos P.;~R{\'{e}}nyi, ``{On random graphs},'' \emph{Publicationes
  Mathematicae}, vol.~6, no.~1, pp. 290--297, 1959.

\bibitem{Barthelemy2003}
M.~Barth{\'{e}}lemy, ``{Crossover from scale-free to spatial networks},''
  \emph{Europhysics Letters (EPL)}, vol.~63, no.~6, pp. 915--921, 2003.

\bibitem{GplusPaper}
J.~McAuley and J.~Leskovec, ``Learning to discover social circles in ego
  networks,'' in \emph{Advances in Neural Information Processing Systems},
  2012, pp. 548--556.

\bibitem{Barbieri2011}
A.~L. Barbieri, G.~de~Arruda, F.~A. Rodrigues, O.~M. Bruno, and
  L.~da~Fontoura~Costa, ``An entropy-based approach to automatic image
  segmentation of satellite images,'' \emph{Physica A: Statistical Mechanics
  and its Applications}, vol. 390, no.~3, pp. 512 -- 518, 2011.

\bibitem{websiteCU}
\BIBentryALTinterwordspacing
G.~Wheeler. (2013) Marketing's new mission: Creating a customer curiosity
  strategy. [Online]. Available: \url{http://www.mycustomer.com}
\BIBentrySTDinterwordspacing

\bibitem{websitePG}
\BIBentryALTinterwordspacing
TEDxZurich. (2017) Pok{\'{e}}mon go is now the biggest mobile game in u.s.
  history. [Online]. Available: \url{http://tedxzurich.com}
\BIBentrySTDinterwordspacing

\end{thebibliography}

\end{document}